\documentclass[british]{elsarticle}
\usepackage{lmodern}
\usepackage[a4paper]{geometry}
\usepackage[T1]{fontenc}
\usepackage{amsmath}
\usepackage{multirow}

\usepackage{tabulary}
\usepackage{amssymb}
\usepackage{bm}
\usepackage{graphicx}
\usepackage{wasysym}
\usepackage{url}
\usepackage{xcolor}
\usepackage{color}
\usepackage{subfig}
\usepackage{enumitem}
\usepackage{mathtools}
\usepackage{threeparttable}
\usepackage{comment}
\usepackage{url}
\usepackage[T1]{fontenc}
\usepackage{versions}
\usepackage{tikz}
\usetikzlibrary{shapes,arrows,matrix,backgrounds,shapes,positioning,petri,topaths}
\usetikzlibrary{chains,calc}
\usepackage{pgfplots}
\pgfplotsset{compat=1.9}
\usepackage{booktabs}
\usepackage{pgf-umlsd}
\usepackage{svg}
\usepackage{siunitx}
\usepackage{amsmath}
\usepackage[ruled,vlined]{algorithm2e}
\usepackage{enumitem}

\usepackage{eqparbox}

\newcommand{\anb}{\textit{AnB}}
\newcommand{\anbx}{\textit{AnBx}}

\newcommand{\jprotocol}{\textit{Typed-Opt-ExecNarr}}
\newcommand{\execnarr}{\textit{ExecNarr}}
\newcommand{\optexecnarr}{\textit{Opt-ExecNarr}}

\usepackage{listings}

\lstdefinelanguage{AnBx}{
	keywords = {Protocol:, Types:, Agent, Number, Function, Certified, SymmetricKey, PublicKey, Untyped, Definitions:, Knowledge:, where, agree, share, Actions:, Goals:, weakly, authenticates, on, secret, between, empty},
	alsoletter=:,
	breaklines=true,
	basicstyle=\footnotesize\ttfamily,%
	keywordstyle=\bfseries,
	commentstyle=\color{green}\itshape,
	morecomment=[l]{\#}
}

\lstdefinelanguage{OFMCAttackTrace}{
	keywords = {Reached, State:,request,state_rB,state_rA,state_rs,state_rIntr,ATTACK, TRACE:},
	alsoletter=:,
	breaklines=true,
	basicstyle=\footnotesize\ttfamily,%
	keywordstyle=\bfseries,
	commentstyle=\color{green}\itshape,
	morecomment=[l]{\#}
}

\newif\ifshowfixes \showfixestrue

\definecolor{pgreen}{rgb}{0,0.5,0}
\usepackage{doi}
\makeatletter
\def\ps@pprintTitle{%
	\let\@oddhead\@empty
	\let\@evenhead\@empty
	\let\@oddfoot\@empty
	\let\@evenfoot\@oddfoot
}
\makeatother
\begin{document}
\title{A Practical Approach to Formal Methods:\\An Eclipse Integrated
Development\\Environment (IDE) for Security Protocols\footnote{This is the extended version of the published paper \cite{anbx-ide-2024} available at \url{https://doi.org/10.3390/electronics13234660}.}}
	\author{Rémi Garcia}
	\ead{r.garcia@tees.ac.uk}
 \author{Paolo Modesti}
	\ead{p.modesti@tees.ac.uk}
	\address{Teesside University, Middlesbrough, United Kingdom}

\begin{abstract}
In order to develop trustworthy distributed systems, verification techniques and formal methods, including lightweight and practical approaches, have been employed to certify the design or implementation of security protocols. Lightweight formal methods offer a more accessible alternative to traditional fully formalised techniques by focusing on simplified models and tool support, making them more applicable in practical settings. The technical advantages of formal verification over manual testing are increasingly recognised in the cybersecurity community. However, applying formal methods, even in their more practical forms, outside highly specialised research settings remains challenging. For practitioners, formal modelling and verification are often too complex and unfamiliar to be used routinely. In this paper, we present an Eclipse Integrated Development Environment for the design, verification, and implementation of security protocols and evaluate its effectiveness, including feedback from users in educational settings. It offers user-friendly assistance in the formalisation process as part of a Model-Driven Development approach. This IDE centres around the Alice \& Bob (AnB) notation, the AnBx Compiler and Code Generator, the OFMC model checker, and the ProVerif cryptographic protocol verifier. We surveyed master-level students on their knowledge prior to completing their projects, and noticed considerable misconceptions. After project completion, we asked them to report on their experience with our IDE. Our findings demonstrate that this contribution is valuable as a workflow aid and helps users grasp essential cybersecurity concepts, even for those with limited knowledge of formal methods or cryptography. Crucially, users reported that the IDE has been an important component to complete their projects, that they would use again in the future, given the opportunity.

\end{abstract}


\maketitle

\section{Introduction}
\label{sec:ideintroduction}

The importance of formal methods in the development of security protocols has become increasingly evident in recent years, as formal verification provides a high degree of confidence in a protocol's security, ensuring that critical flaws are identified early in the protocol design phase, thereby reducing the risk of exploitation in real-world deployments.

Serious vulnerabilities have been uncovered in several protocols that constitute the backbone of the digital infrastructure allowing individuals and organisations to perform their online activities.

For example, a critical design-level vulnerability such as the KRACK attack \cite{DBLP:conf/ccs/VanhoefP17} on the WPA2 Wi-Fi standard allowed for decryption, packet replay, TCP connection hijacking, and HTTP content injection on nearly all devices using the standard. TLS, one of the most ubiquitous protocols, has also exposed serious vulnerabilities, leading to various attacks, including Heartbleed \cite{heartbleed2014}, Freak \cite{Fogel2016}, and downgrade attacks \cite{Alashwali2018}.

This situation has driven the research and development community to place a greater emphasis on rigorous verification methods. Several authors have now formally verified TLS 1.3 \cite{Cremers2016,Bhargavan2017,blanchet2018composition}, and similar notable work has been undertaken for other protocols, including Signal \cite{DBLP:journals/joc/Cohn-GordonCDGS20}.

These efforts include the use of automated verifiers such as ProVerif \cite{blanchet2001efficient}, Tamarin \cite{DBLP:conf/cav/MeierSCB13}, and OFMC \cite{basin2005ofmc}, among others. The application of these tools has enabled security specialists to design robust protocols during the early stages of development, rather than relying on a reactive approach of addressing vulnerabilities post-deployment.

Despite these advances, attacks continue to exploit vulnerabilities in software implementations \cite{Dark2015c}. Those include buffer overflows in OpenSSL with the HeartBleed bug \cite{heartbleed2014} leaking sensitive information, and data confusion exploits on GnuTLS \cite{CVE-2020-13777}, where a server can be tricked into using wrong data as an encryption key. Those attacks are immediately applicable to a wide range of targets, as they leverage vulnerabilities in security libraries. In such cases, missing defensive run-time checks at the implementation level enabled those attacks, even though such checks were implied, even implicitly, in the design phase, but failed to materialise or work in the concrete code. To alleviate those issues, there is now ample consensus among experts that the application of formal methods for security should be more widespread \cite{DBLP:conf/fmics/GaravelBP20,DBLP:journals/fac/KulikDLMSTW22}. Formal-backed code generation in a convenient environment would bridge the gap between formal design and implementation. It would also show examples of necessary implementation checks that reflect the design intent and prevent misconceptions or oversights.\\

Prior to the implementation of a protocol, its design requires considerable effort, involving extensive expertise and verification.

First of all, a fundamental security paradigm that must be taken into account is the Dolev-Yao attacker model \cite{dolev83ieee}, which defines the assumptions about the attacker's capabilities. Specifically, it assumes that the adversary has total control over the network, can intercept and modify data in transit, and can perform composition and decomposition operations using the same rules as honest agents. This paradigm has practical applications in the development of the aforementioned verification tools.

Having a suite of tools for the formal modelling and analysis of security protocols to support software engineers is essential. However, when multiple tools need to be used in coordination, this creates a significant complexity barrier for users due to their complexity, unfamiliar interfaces, and difficulties in the interpretation of results \cite{DBLP:journals/fac/KulikDLMSTW22}.

Our research aims to make this initial design step more accessible for non-experts and more convenient for experts, so we can encourage the use of formal methods for security protocols and avoid the high costs associated with late vulnerability detection \cite{sommerville2013software} or the potentially catastrophic consequences of deploying vulnerable protocols at large scale. In general, a lack of control over the specification and verification process could, at best, lead to misunderstandings. A beginner would benefit from being guided by warnings, error messages, and a simple push-button solution, while an expert could leverage more advanced features for quality control, all while enjoying increased productivity. Thus, once a design is established, it could be thoroughly verified without the need to manage the complexity of command-line interactions.

This design effort must also be supported beyond dedicated tools for a specific modelling language. When outlining the skeleton of a protocol, one should be able to reason about it without having to rely on a lower-level language, which would be time-consuming. Therefore, we advocate for a very high-level specification language, accompanied by a suite of tools built around it. 

The research community \cite{bugliesi2008lbs,avalle2014formal} supports the notion of high-level abstractions, and the 2020 Expert Survey on Formal Methods \cite{DBLP:conf/fmics/GaravelBP20} highlights that the most reported limiting factors for the adoption of formal methods are the lack of training, tool limitations and insufficient maintenance, and the lack of integration of formal methods into the industrial design life cycle. 

The educational barrier to adopting formal methods is mentioned in one comment: ``Education in formal methods frightens off students and puts them off for life rather than showing potential benefits.''. Furthermore, integration of formal methods tools in general, and in IDEs specifically, has been recommended by many experts \cite{DBLP:conf/fmics/GaravelBP20,DBLP:journals/fac/KulikDLMSTW22} and identified as one of the most important measures to overcome the barriers for industry adoption \cite{Davis2013}.

Tool integration can be an important way to enable education and training, as this is the other major barrier identified in relevant studies \cite{DBLP:conf/fmics/GaravelBP20,DBLP:journals/fac/KulikDLMSTW22,Davis2013}. Therefore, our approach to tool integration also aims to facilitate training, thereby lowering the barriers and yielding positive outcomes for learners, both in terms of reducing error rates and increasing confidence, as identified by Skevoulis \textit{et al.} \cite{Skevoulis2006}. This study suggests that a lack of integrated tools constitutes a significant challenge for formal methods adoption, highlighting the need for more intuitive educational approaches.

Other studies \cite{Scheurer2000,Pomorova2016} also recognise the pedagogic value of integrated formal methods tools to improve comprehension and engagement.

\subsection*{Research contribution}

The main objectives of our work focus on the verification of symbolic models for cryptographic abstractions \cite{blanchet2001efficient,basin2005ofmc,DBLP:conf/cav/MeierSCB13}, alongside the automatic generation of security protocols at both the specification and implementation levels \cite{DBLP:journals/ijsse/AvallePPS11,anbx2015,sps2015}. This approach not only reduces the complexity of the tasks but also promotes the broader adoption of formal methods in the development of secure protocols.

In order to encourage users to leverage formal methods, we propose an approach based on the following key components:

\begin{itemize}
	\item A simple and intuitive language for lightweight \cite{wing1996lightweight,DBLP:conf/enase/ZamanskySRHB18} formal specification of security protocols. This addresses the problem of complexity in formal methods by helping users familiarise themselves with system design from a high-level perspective, before delving into verification. The expected benefit is that users, especially non-experts, can engage with formal methods more easily and develop a better understanding of protocol design without being overwhelmed by technical details \cite{Brucker2024}.
	
	\item \textit{A Model-Driven Development} (MDD) strategy for the automatic generation of programs from formally verified abstract models. This eliminates the need for users to prototype at multiple abstraction levels, ensuring consistent and rigorous translation from model to implementation. The benefit is a reduction in development time and errors, as the MDD approach streamlines the workflow, allowing users to focus on design correctness rather than manual coding at each layer \cite{DBLP:series/synthesis/2017Brambilla,DBLP:journals/infsof/NguyenKKT15}.
	
	\item An \textit{Integrated Development Environment} (IDE) with push-button integration of existing tools and real-time modelling/verification feedback. This addresses the issue of tool complexity and integration. By providing a seamless environment for developers, it automates significant parts of the workflow and offers real-time feedback, thus improving both the clarity of the specification process and control over the verification stage. 
\end{itemize}

In this paper, we focus on the latter element, the environment, which is supporting the overall approach, leveraging the two other components.

The \emph{\anbx{} IDE}\footnote{Available at \url{https://www.dais.unive.it/~modesti/anbx/ide/}} is an Integrated Development Environment for the design, verification, and implementation of security protocols. It has been actively maintained and enhanced since its first prototype released in 2017 \cite{anbx-ide2017} and aims to build on this by offering a workflow that benefits both novices and experts alike.

It is developed as a plug-in using XText \cite{Bettini2016,xtext2024}, a framework for creating software languages, including modelling languages, and leverages existing tools for specification, code generation, and verification of security protocols.

The purpose of the \emph{\anbx{} IDE} is to provide an all-in-one solution for users, offering support for high-level protocol specification, automatic code generation, and seamless integration with verification tools. Specifically, it facilitates:
\begin{itemize}
	\item \textit{Specification}: The IDE allows users to specify protocols in the simple Alice \& Bob (\anb{}) notation \cite{DBLP:conf/IEEEares/Modersheim09} or its extension \anbx{} \cite{jisa_2016}, a high-level abstraction for defining cryptographic protocols.
	\item \textit{Code generation}: The \emph{\anbx{} Compiler and Code Generator} \cite{anbx2015} automatically generates ProVerif specifications and Java implementations, among other targets, from the \anb{} and \anbx{} models.
	\item \textit{Verification}: The IDE integrates the symbolic model checker OFMC \cite{basin2005ofmc} and the cryptographic protocol verifier ProVerif \cite{blanchet2001efficient} to allow for automated formal verification of protocols. This integration provides users with a smooth compilation and verification workflow.
\end{itemize}

At the time of writing, the IDE supports the \anb{} notation of OFMC \cite{DBLP:conf/IEEEares/Modersheim09}, the applied-pi calculus utilised by ProVerif \cite{blanchet:hal-03366962}, and the \anbx{} language used to specify security protocols \cite{jisa_2016}. The integration of these verification tools within the IDE facilitates a streamlined design and verification process. This aligns with observations made by Kulik \textit{et al.} \cite{DBLP:journals/fac/KulikDLMSTW22}, who highlight that incorporating verification tools directly into development environments promotes the adoption of formal methods in industry, as evidenced by the growing use of static code analysers \cite{8312264}.

The IDE, along with the interaction with the back-end tools, includes
many features meant to help programmers to increase their productivity,
like syntax highlighting, code completion, code navigation, and quick
fixes. After verification, the IDE offers visual interpretation of the results for clear and immediate user understanding. We put a particular emphasis on the workflow aspects of formal verification. This IDE automates many intermediate steps and key aspects of this effort, such as scheduling, task management, and result visualisation. Those aspects combine usability and productivity advantages for users at all levels. 

This contribution represents a complete overhaul of the preliminary version of the tool introduced in 2017 \cite{anbx-ide2017}. The earlier version only supported \anb{}-style specifications, with minimal workflow automation and no mechanisms for handling verification results or managing tasks effectively. In this revised version, we have added numerous features, observing students' work and feedback to enhance usability, along with a survey to assess the usefulness of the tooling suite.\\

In general, incorporating formal methods into a computer science curriculum is beneficial not only for producing robust software and hardware but also for fostering a deeper understanding of the underlying logical principles \cite{4283790,Skevoulis2006}.
For example, Broy \textit{et al.} \cite{10.1145/3670795}, argue for the teaching of formal methods, emphasising that it provides a foundation for rigorous reasoning about specifications. In line with their argument that basic formalism should be taught to all students, while specialised formal methods are reserved for specific fields, we think, supported by feedback from our students, that our IDE could serve as a valuable tool to facilitate the integration of formal methods into computer science courses.

\subsection*{Outline of the Paper}

We begin by detailing the motivation behind this approach in Section \ref{sec:idemotivation}. Next, in Section \ref{sec:idebackground}, we introduce the specification languages and the back-end tools and in Section \ref{sec:methodology} we present our methodology. Section \ref{sec:idecontribution} presents the IDE’s construction and features and Sections \ref{sec:ideevaluation} and \ref{sec:iderelatedwork} provide our evaluation and discuss related work. Finally, Section \ref{sec:ideconclusion} concludes the paper and outlines potential future work related to our approach.

\section{Motivation}
\label{sec:idemotivation}

In this section, we discuss the challenges in usability and adoption of formal methods, the need for lightweight and practical formal methods, as well as the necessity for Integrated Development Environments that can support the utilisation of associated tools. Moreover, we consider the cryptographic misconceptions observed in a population of cybersecurity students, which further reinforces the need to provide supportive development tools that can help overcome such misconceptions.

\subsection{Challenges in usability and adoption of Formal Methods}
\label{subsec:idechallenges}

Formal methods, while technically successful, have not become a universally adopted solution. Similar to programming languages, specialised methods have emerged for specific fields to avoid the complexities associated with a  ``one size fits all'' approach. A study on the barriers to the industrial adoption of formal methods \cite{Davis2013} suggests that education and tooling are the two most important factors, with highly specialised experts being needed to operate tools that are often neither integrated nor compatible with development tools. The timeline issue is also raised, with formal analysis being perceived as too long a process to be applicable. Demonstrating the practicality of such methods, both for potential and current professional users, could bring formalisation to the foreground, as discussed in \cite{DBLP:journals/fac/KulikDLMSTW22,DBLP:journals/ese/GleirscherM20}. 

To aid in this matter, lightweight formal methods \cite{wing1996lightweight,DBLP:conf/enase/ZamanskySRHB18} are commonly endorsed in the literature and are limited in terms of the languages that can be automatically analysed, contrasting with earlier approaches such as Z \cite{DBLP:books/daglib/0068766}. These methods are abstract enough to remain practical while focusing on the essential aspects that need to be formalised. In this context, our interest lies specifically in security protocols rather than the underlying machines on which they operate.

The lightweight approach is employed in this paper, starting from the observation that many developers are largely unfamiliar with formalisation concepts. For formal methods to be effectively implemented, a fundamental understanding is essential. For instance, the \emph{constructivist learning theory} \cite{wadsworth1996piaget,bruner2009process} suggests that new knowledge is constructed through a combination of prior knowledge and current experience. 
Surveys on the practicality and deployment of formal methods \cite{DBLP:journals/fac/KulikDLMSTW22,DBLP:conf/fmics/GaravelBP20} state that this integration should come to software-build toolchains and not for the purposes of formalism only. This cross-integration approach is also advocated by Davis \cite{Davis2013} \textit{et al.}

To facilitate the learning process, our objective is to eliminate obstacles to experimentation. Users should not struggle with setup procedures before they can begin specifying and verifying protocols. By enabling users to achieve results with minimal time and effort, we aim to prevent discouragement and promote continued engagement in the formalisation process.

To address these complexity challenges, an intuitive language in the Alice \& Bob style is essential to enable learners to understand fundamental concepts in formal methods and security protocols \cite{Brucker2024}, and the relevance of \anbx{} has been emphasised by an external study in \cite{DBLP:conf/etfa/RaimondoMBP23}, particularly to support model-driven security analysis. Brucker \textit{et al.} \cite{Brucker2024} mention the differences in teaching experience between  \anb{} with OFMC, which successfully enabled undergraduates to grasp Dolev-Yao formalism, and more specialised languages and frameworks as Isabelle/HOL \cite{DBLP:journals/fac/PaulsonNW19}, which proved to be too complex even for postgraduate students. However, despite being considerably simpler than other languages (e.g., the applied-pi calculus used by ProVerif), \anbx{}/\anb{} can still present technical challenges for beginners. For instance, the distinction between symmetric and asymmetric encryption is not always immediately clear. Furthermore, in the context of asymmetric encryption, the correct usage of keys for different purposes (i.e., encryption versus signature) is crucial, and it is essential that key material is not confused \cite{abadi1994pep}. In \anbx{}, this distinction is reinforced by providing distinct predefined functions: \emph{pk(.)} for keys used for asymmetric encryption and \emph{sk(.)} for keys used for digital signature. 

By combining an intuitive notation with supportive tooling, the amount of experience one can gain with cybersecurity concepts can be significantly increased, with much less difficulty than would be observed without such support. Lowering the adoption barrier for formal verification not only enables users to become proficient in the formal aspects of security but also helps them develop a deeper understanding of more general security-related concepts. In fact, sketching ideas in a high-level notation and receiving immediate feedback from formal verification creates a rapid learning loop that encourages experimentation. This approach ensures at least a high-level understanding of how the underlying primitives operate.
\\

The core usability challenges and limits in the widespread adoption of formal methods for security, in light of the 2020 Expert Survey on Formal Methods \cite{DBLP:conf/fmics/GaravelBP20}, the study on the barriers to the industrial adoption of formal methods \cite{Davis2013}, and other work on formal methods usability issues and practical challenges
\cite{DBLP:journals/corr/LeinoW14,DBLP:journals/fac/KulikDLMSTW22}, can be summarised as follows:

\begin{enumerate}
    \item \textit{Complexity:} The inherent complexity of these methods often requires users to possess a good understanding of mathematical concepts and logic, resulting in a high learning curve that can be deterring for practitioners without a formal background, or knowledge of a specific verification framework.
    \item \textit{Limited Tool Integration:} The limited availability of well-integrated tools can create inefficiencies in workflow, hindering effective utilisation of formal methods. In particular, a majority of criticisms pointed out that research is not enough oriented towards the consolidation of existing results to make them available to a wider audience.
    \item \textit{Unfamiliar Interfaces:} Specialised, unusual, and inconsistent user interfaces across different tools may lead to confusion, making it challenging for users to navigate various applications.
    \item \textit{Interpretability of Results:} The interpretability of results is often compromised due to complex output formats and cryptic error messages, making it difficult for users to derive actionable insights.
    \item \textit{Scalability Issues:} Many formal verification tools struggle to handle complex systems efficiently, resulting in long processing times and significant resource consumption.
    \item \textit{Limited Documentation:} Limited or hard to find documentation and community support exacerbate these challenges, leaving users without sufficient guidance.
\end{enumerate}

\subsection{The benefits of Integrated Development Environments}

Before implementing an IDE for \anbx{}, we conducted tutorial sessions in which students used command-line tools for modelling, verification, and code generation, alongside simple text editors. However, the setup of these tools and managing the complexity of the toolchain proved to be a deterrent and a distraction for some students, particularly those with lower skills or motivation. We also considered a simpler syntax-highlighting solution in Notepad++, but this lacked critical features, such as real-time validation feedback from the editor and flexible, user-friendly tool parameter selection.

Although a script-based approach could have been applied for security protocols, we chose a different direction in this case. This decision was motivated not only by some users’ preference for graphical user interfaces (GUIs) over command-line tools \cite{Unwin1999}, but also by the complexity of integrating a toolchain with multiple heterogeneous tools, which requires a level of scrutiny beyond that of a standard, well-tested programming language compilation chain. While scripts can be highly effective for experts, they tend to be less accessible for beginners.

Given these considerations, we opted to support users with an IDE, as Integrated Development Environments not only enhance productivity but also lower the adoption barrier for new methodologies and technologies.

To leverage languages and back-end tools effectively, an integration effort can foster openness and interest among developers in adopting such tools. An IDE can greatly simplify the setup and configuration of the environment. We align with the perspective of Tabassum \textit{et al.} \cite{DBLP:conf/soups/TabassumWL17}, believing that an IDE can facilitate secure coding by providing instant security warnings, detailed explanations, and more. The Kulik \textit{et al.} \cite{DBLP:journals/fac/KulikDLMSTW22} and the Davis \cite{Davis2013} \textit{et al.} surveys report that there is a general consensus on advocating for the improvement of automated formal verification tools, to ease the entry into the formal methods world. Direct integration into an existing IDE is mentioned as a step in this direction, and we aim to participate in this effort.
There is also significant empirical evidence supporting this claim; widely used IDEs such as Eclipse, Visual Studio and the IntelliJ suite demonstrate that developers appreciate the functionality they provide. The choice of Eclipse here is motivated by its extensive and continued support for XText.

On a broader scale, Kuusinen  \cite{DBLP:conf/profes/Kuusinen15} conducted a survey involving 45 developers from 21 countries to determine their ideal features in an IDE. In addition to being efficient and flexible, developers expressed a desire for the IDE to be informative and intuitive to use. They indicated that it should support a cohesive development workflow, offering control and clear visibility of both the code and the overall development situation. This is precisely what we aim to provide through features such as editing, visualisation, and task management.

\subsection{Identification of potential cryptographic misconceptions}
\label{subsec:survey-misconceptions}

Initially, we sought to assess students' knowledge of applied cryptography to better tailor teaching and learning strategies in our university courses. Understanding students' misconceptions and their baseline knowledge is vital for designing effective educational interventions, particularly in a specialised field like cybersecurity. The results also enabled us to understand how an IDE for security protocols could support users in avoiding the misuse of cryptography. 

To achieve this, we developed a survey consisting of 24 multiple-choice questions. For each question, there were four options, one of which was correct, with the exception of a few questions that offered only three options, as the nature of those questions allowed for just three plausible choices. The survey includes some questions about general cybersecurity knowledge but focuses on applied cryptography. It covers concepts like authentication, confidentiality, types of communication channels, symmetric and asymmetric encryption, message authentication codes and hashing functions, 

The questionnaire was administered between 2021 and 2024 to four groups of students enrolled in the MSc Cybersecurity course at Teesside University in the United Kingdom, during lab activities, and they were not allowed to consult any sources of information while answering the questions that were presented in random order.

In total, we surveyed 59 students, 48 male and 11 female. The backgrounds of the participants were diverse, though none had an undergraduate degree specifically in Cybersecurity. Most of of the participants had studied computing related degree (Computer Science, Networking and Information Technology), a few individuals had Engineering degrees in fields unrelated to computing. With a few exceptions, the students completed their undergraduate studies abroad, with Nigeria and India being the most represented countries. 25 participants took the survey at the beginning of their postgraduate studies, while the rest completed it after a year.

To further understand the technical competencies of the students and identify areas requiring further development, an initial skills self-assessment was also conducted. This enquiry considered proficiency in key areas, including networking, system administration, cybersecurity fundamentals, and programming.  

Overall, for the purpose of this research, the sample is representative of a population of different educational background and level of expertise. A significant number of students (71\%) rated their networking skills as \textit{Strong}, while 57\% felt the same about their Windows system administration capabilities. In contrast, only 21\% of students considered themselves \textit{Strong} in Linux system administration, highlighting a major area of concern where 50\% reported feeling \textit{Weak}. Ethical hacking skills showed mixed results, with 14\% rating themselves as \textit{Very Strong}, 36\% as \textit{Strong}, 43\% as \textit{Average}, and 7\% as \textit{Weak}. Finally, in scripting and programming, only 14\% of students felt \textit{Strong}, while 64\% rated their skills as \textit{Weak}, indicating a significant need for improvement in this area. Their limited mastery of programming and security subjects makes them ideal candidates for this study: without a strong foundational understanding, they are more prone to misconceptions in cryptography and would likely face increased confusion and disengagement without adequate supporting tools.
\\

The results of the applied cryptography survey allowed us to identify the most prominent cryptography misconceptions, as a strong understanding of applied cryptography at the symbolic level, used by the verification tools in this study, is crucial for modelling protocols and interpreting the results of verification tools.
The identified misconceptions are reported below:

\textbf{Symmetric vs Asymmetric Encryption}
Over a third (39\%) did not know that the same key is used for both encryption and decryption in symmetric cryptosystems, and 22\% confused it with other systems, stating that different independently generated keys are used. 

22\% of the respondents incorrectly claimed that the main disadvantage of symmetric encryption is algorithm complexity, while 31\% believed it is less secure. The core issue lies in securely sharing the secret key, which only 48\% understood.

The distinction between symmetric and asymmetric cryptography is often unclear; 29\% mistakenly believed that public key cryptography is symmetric, and 22\% that it is both symmetric and asymmetric.

\textbf{Usage of Encryption}
Consequently, less than half (44\%) correctly identified that asymmetric cryptography can be used for encryption, digital signatures, and integrity verification. Misunderstandings persist, with 39\% unaware of who should use the private key for decryption and 25\% mistakenly thinking it should be used by both sender and receiver.

Only 17\% correctly identified that the recipient's public key should be used to establish a confidential channel, while 40\% thought it should be the recipient's private key. This indicates that over 80\% are unfamiliar with basic encryption practices.

Further misunderstandings arise regarding signature keys; more than half suggested transmitting a private key, either encrypting with the private key (27\%) or public key (25\%). Many did not realise this would compromise the secrecy of the private key. The correct option, using the private key without sending it directly, was chosen by only 29\%.

\textbf{Hash and Message Authentication Codes (MACs)}
Definitions of concepts like Hash functions and MACs are often misunderstood. A staggering 78\% incorrectly believed that a hash function's output is non-deterministic. Only 40\% recognised that hash functions compress input values, work on arbitrary-length input, and produce fixed-length output.

Regarding HMAC, 54\% provided the correct definition, while 22\% incorrectly associated it with the Diffie-Hellman key derivation. Other misconceptions included defining HMAC as high-integrity medium access control (17\%) or as a self-healing algorithm for stream ciphers (7\%).

Awareness of potential collisions is also low; 36\% stated that hash functions are unrelated to collision resistance, and only 29\% knew that a MAC could correspond to multiple messages. Meanwhile, 44\% assumed a one-to-one relationship, while 19\% mistakenly thought multiple MACs could correspond to a single message.

\textbf{Authentication and Confidentiality}
The cryptographic concept of authentication is widely misunderstood. While some respondents recognised it can be achieved through something known (24\%), inherent (15\%), or possessed (8\%), only 53\% identified all options as valid. Regarding its purpose, 36\% mistakenly thought it did not involve agreeing on message content, and 76\% viewed it as an access control mechanism.

Many (32\%) were unaware that faulty authentication could lead to Man-in-the-Middle attacks. Similarly, solely 39\% correctly answered that mutual authentication could prevent all three of Man-in-the-Middle, spoofing, and replay attacks. For scenarios benefiting from unilateral authentication, only 24\% recognised its relevance with pseudonyms.

Best practices for security are poorly understood; only 27\% acknowledged that switching off a device when unused does not adequately maintain confidentiality. Students must learn that confidentiality relies on tailored security measures and network security under Dolev-Yao assumptions.

\textbf{General Knowledge}
We also posed general questions about existing technologies. A substantial 84\% correctly indicated that a secure channel must be both authentic and confidential, and 76\% identified “an internet scam convincing users to provide confidential information” as phishing.

However, the proportion of correct answers diminishes with more specific topics. Only 58\% knew that PKI stands for Public Key Infrastructure, while 57\% recognised that a certificate from a Certificate Authority guarantees a public key's ownership. A mere 37\% could identify Diffie-Hellman as a method for generating a one-time session key, and just 31\% knew that TLS secures HTTP connections, often confused with SSH (34\%).
\\

Although the diverse educational backgrounds of participants may explain some incorrect responses, significant misconceptions still emerged. The lack of formal training in cybersecurity in their undergraduate studies and the different levels of expertise, particularly among students at the beginning of their postgraduate studies, likely contributed to these misunderstandings. Those results align with other studies presented by Lindmeier \textit{et al.} \cite{DBLP:conf/wipsce/LindmeierM20} or Geels \cite{DBLP:conf/nordichi/Geels24}, where fundamental beginner mistakes on definitions, cryptography usage, and network assumptions happen regularly. Cryptography misuse has also been observed in \cite{DBLP:journals/tr/BragaDALV19}, browsing programming forums.

Overall, the survey results highlighted several misconceptions among participants that have significant implications for the development of an IDE for security protocols using symbolic models in the Dolev-Yao style, such as the Alice \& Bob notation.

A notable number of students struggled to differentiate between symmetric and asymmetric encryption, with many unaware that the same key is used for both encryption and decryption in symmetric systems. This lack of understanding can hinder their ability to model protocols accurately. 

Moreover, misconceptions about the correct applications of asymmetric cryptography, where participants failed to recognise agents' roles in encryption, digital signatures, and integrity verification, along with confusion surrounding private keys, many incorrectly believing they should be shared, suggest a need for the IDE to incorporate elements that address the misuse of such concepts.

Given these identified misconceptions, it is clear that there is considerable potential for progress in an appropriate environment. The 2020 expert survey \cite{DBLP:conf/fmics/GaravelBP20} supports this sentiment: they largely consider that while formal methods bring success on a technical level, there remains considerable unexploited potential for these methods to become more widespread among practitioners. In particular, these experts point out that both education and software tools are lacking to support broader adoption of formal methods.

\section{Background}
\label{sec:idebackground}

In this section, we describe the XText framework, which serves as the foundation for our Integrated Development Environment. We outline the input languages and back-end tools currently supported by the IDE, highlighting its integration of the \anbx{} Compiler and Code Generator, the model checker OFMC, and the cryptographic protocol verifier ProVerif. This integration enables users to effectively specify, verify, and implement security protocols, streamlining the development process.

\subsection{XText}

XText \cite{Bettini2016,xtext2024} is a popular workbench for the development of Domain-Specific Languages (DSLs), which is part of the Eclipse ecosystem. It facilitates the automatic generation of essential components such as parsers, validators, formatters, and highlighters, supporting the customisation effort for any given language by providing a robust existing architecture. In Xtext, grammars are expressed in Extended Backus-Naur Form (EBNF), while other components are specified using XTend, a language based on Java that aims to be more concise, readable, and expressive. Additionally, a Model Workflow Engine (MWE) file orchestrates the generation of concrete artefacts from these specifications, allowing them to be utilised as plug-in components. The use of the XText framework for Domain-Specific Languages, such as \anbx{}, presents a significant opportunity for integrated formal methods, as discussed in \cite{DBLP:journals/csur/GleirscherFW20}. This framework has been effectively applied in various works \cite{DBLP:conf/ifm/GlabbeekHW18,DBLP:conf/se/RungeSCTKW21,DBLP:conf/facs2/FaresBF24}, including for common equipment like Arduino, with Asm2C++ \cite{DBLP:conf/nfm/BonfantiCGM17}, translating formal Abstract State Machines to targeted C++.

\subsection{\anbx{} Language}

\begin{figure}[t]
\begin{lstlisting}
Protocol: Example_AnBx
Types:
    Agent A,B;
    Certified A,B;
    Number Msg;
    SymmetricKey K;
    Function [Agent,Number -> Number] log
Knowledge:
    A: A,B,log;
    B: B,A,log
Actions:
    A -> B, @(A|B|B): K
    B -> A: {|Msg|}K
    A -> B: {|hash(Msg),log(A,Msg)|}K
Goals:
    K secret between A,B
    Msg secret between A,B
    A authenticates B on Msg
    B authenticates A on K,Msg
\end{lstlisting}

\caption{\label{fig:AnBx-Protocol-Example}\anbx{} Protocol Example}
\end{figure}
\begin{figure}[t]
\begin{lstlisting}
Protocol: Example_AnB
Types:
    Agent A,B;      # Certified
    Number Msg,Nonce;
    Symmetric_key K;
    Function pk,sk; # [Agent] -> PublicKey
    Function hash;  # [Untyped] -> Number
    Function log    # [Agent,Number] -> Number
Knowledge:
    A: A,B,pk,sk,inv(pk(A)),inv(sk(A)),hash,log;
    B: A,B,pk,sk,inv(pk(B)),hash,log
Actions:
    A -> B: A
    B -> A: {Nonce,B}pk(A)
    A -> B: {{Nonce,B,K}inv(sk(A))}pk(B)
    B -> A: {|Msg|}K
    A -> B: {|hash(Msg),log(A,Msg)|}K
Goals:
    K secret between A,B
    Msg secret between A,B
    A authenticates B on Msg
    B authenticates A on K,Msg
    inv(pk(A)) secret between A
    inv(sk(A)) secret between A
\end{lstlisting}

\caption{\label{fig:AnB-Protocol-Example}\anb{} Protocol Example}
\end{figure}

The \anbx{} language is an extension of \anb{} \cite{DBLP:conf/IEEEares/Modersheim09}, and formally defined in \cite{jisa_2016}. On top of type signatures, \anbx{} introduces flexible channel notations as abstractions for communication. Those can
provide different authenticity and confidentiality guarantees for
message transmission, including a novel notion of \emph{forwarding} channels, enforcing security guarantees from the message originator to the final recipient along a chain of intermediate agents.

Figure \ref{fig:AnBx-Protocol-Example} shows an example protocol
in which two agents want to exchange securely a message \texttt{Msg},
using a freshly generated symmetric key \texttt{K}, i.e., a key that
is different for each protocol run. If \texttt{K} is compromised,
neither previous nor subsequent exchanges will be compromised,
but only the current one. This is similar to what happens in TLS, where
a symmetric session key is established (using asymmetric encryption)
at the beginning of the protocol. It should also be noted that this
setting is also more efficient, as symmetric encryption is notoriously faster than the asymmetric one. 

The \texttt{Types} section includes declarations of identifiers of
different types and functions declarations, while the section \texttt{Knowledge}
denotes the initial knowledge of each agent. Optional sections,
\texttt{Definitions} and \texttt{Equations}, can be used respectively to specify macros with parameters or equational theories.
In the \texttt{Actions} section, the action \texttt{A -> B,@(A|B|B):K}
means that the key \texttt{K} is generated by \texttt{A} and sent
on a secure channel to \texttt{B}. The notation \texttt{@(A|B|B)}denotes
the properties of the channel: the message originates from
\texttt{A}, it is freshly generated (\texttt{@}), verifiable by \texttt{B},
and secret for \texttt{B}. How the channel is implemented is delegated
to the compiler. The designer can select between different options
or simply use the default one, without having to take charge of low-level implementation details.
A translation to \anb{} is shown in Figure \ref{fig:AnB-Protocol-Example}:
in this case, the channel is implemented using a challenge-response
technique, where \texttt{B} freshly generates a \texttt{Nonce} (the
challenge), encrypted with \texttt{pk(A)}, the public key of \texttt{A}
along with the sender name (\texttt{\{.\}} denotes the asymmetric
encryption). This guarantees that only \texttt{A} would be able to
decrypt the incoming message.

The response, along with the challenge, includes the symmetric key
\texttt{K}. The response is digitally signed with \texttt{inv(sk(A))},
the private key of \texttt{A} and then encrypted with \texttt{pk(B)},
the public key of \texttt{B}. This allows \texttt{B} to verify
the origin of the message and that \texttt{K} is known only by \texttt{A} and \texttt{B}.

It should be noted that in \anbx{}, we abstract from these cryptographic
details, and we simply denote the capacity of \texttt{A} and \texttt{B}
to encrypt and digitally sign using a Public Key Infrastructure (PKI)
with the keyword \texttt{Certified}. This reflects the customary practice
of a Certification Authority to endorse the public keys of agents, usually
issuing X.509 certificates, allowing every agent to verify the identity
associated with a specific public key. Moreover, in \anbx{}, keys
for encryption and for signature are distinguished by using two different
symbolic functions, \texttt{pk} and \texttt{sk} respectively.

Once the symmetric key \texttt{K} is shared securely between \texttt{A}
and \texttt{B}, then \texttt{B} can send the payload \texttt{Msg}
secretly (\texttt{\{|.|\}} denotes the symmetric encryption). Finally,
\texttt{A} acknowledges receipt by replying with a digest of \texttt{Msg}
computed with the predefined \anbx{} \texttt{hash} function, and with a value computed with the \texttt{log} function.

The section \texttt{Goals} denotes the security properties that the
protocol is meant to convey. They can also be translated into low
level goals suitable for verification with various tools. Supported
goals are: 1) \textit{Weak Authentication} goals have the form \texttt{B
	weakly authenticates A on Msg} and are defined in terms of non-injective
agreement \cite{Lowe97}; 2) \textit{Authentication} goals have the
form \texttt{B authenticates A on Msg} and are defined in terms of
injective agreement on the runs of the protocol, assessing the freshness
of the exchange; 3) \textit{Secrecy} goals have the form \texttt{Msg
	secret between A1,...,An} and are intended to specify which agents
are entitled to learn the message \texttt{Msg} at the end of a protocol
run.

In the example protocol (Figure \ref{fig:AnBx-Protocol-Example}),
the desirable goals are the secrecy of the symmetric key \texttt{K}
and of the payload \texttt{Msg} that should be known only by \texttt{A}
and \texttt{B}. There are also authentication goals: \texttt{B} should
be able to verify that \texttt{K} originates from \texttt{A} and the
key is freshly generated. Finally, two goals express the mutual authentication
between \texttt{A} and \texttt{B} regarding \texttt{Msg}, including
the freshness of the message. In summary, this protocol allows two
agents to securely exchange a message, with guarantees about the origin and the freshness of the message.

\subsection{\anbx{} Compiler and Code Generator}

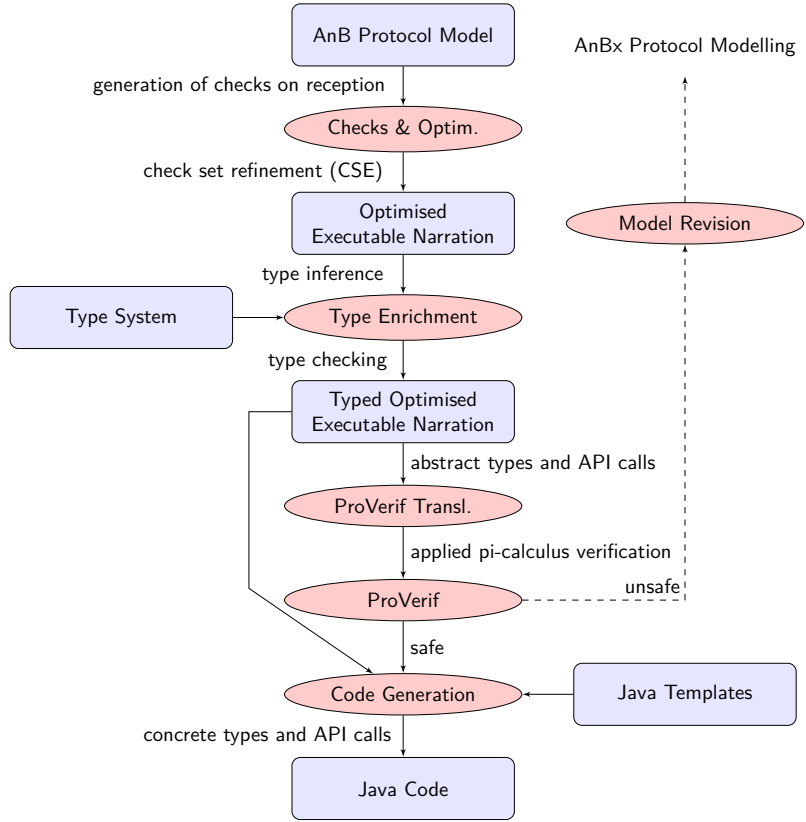
\begin{figure}[!h]
\begin{centering}
\resizebox{110mm}{!}{

\tikzstyle{decision} = [diamond, draw, fill=blue!10,text width=4.75em, text badly centered, node distance=1.5cm, inner sep=0pt,font=\sffamily]
\tikzstyle{block} = [rectangle, draw, fill=blue!10,text width=10em, text centered, rounded corners, minimum height=3em, minimum width=3cm,font=\sffamily]
\tikzstyle{line} = [draw, -latex',font=\sffamily]
\tikzstyle{cloud} = [draw, ellipse,fill=red!20, node distance=1.6cm, minimum height=2em, minimum width=4cm, font=\sffamily]

\begin{tikzpicture}[node distance = 1.6cm, auto]
    \node [block] (anb) {AnB Protocol Model};
    \node [cloud, below of=anb] (anb2execnarr) {Checks \& Optim.};
    \node [block, below of=anb2execnarr] (execnarr) {Optimised\\Executable Narration};
	\node [cloud, below of=execnarr] (typing) {Type Enrichment};
    \node [block, left of=typing,node distance=4.75cm] (typesystem) {Type System};
	\node [block, below of=typing] (typedexecnarr) {Typed Optimised Executable Narration};
	\node [cloud, below of=typedexecnarr] (execnarr2proverif) {ProVerif Transl.};
    \node [cloud, below of=execnarr2proverif] (proverif) {ProVerif};
    \node [cloud, right of=execnarr,node distance=4.75cm] (revise) {Model Revision};
	\node [cloud, below of=proverif] (javacodegen) {Code Generation};
    \node [block, below of=javacodegen] (javacode) {Java Code};
    \node [block, right of=javacodegen, node distance=4.75cm] (javatemplates) {Java Templates};

    \path [line] (anb) -- node[xshift=-5.35cm] {generation of checks on reception} (anb2execnarr);
	\path [line] (anb2execnarr) -- node[xshift=-4.5cm] {check set refinement (CSE)}(execnarr);
	\path [line] (execnarr) -- node [xshift=-2.5cm] {type inference} (typing);
	\path [line] (typesystem) -- (typing);
	\path [line] (typing) -- node [xshift=-2.4cm] {type checking} (typedexecnarr);
	\path [line] (typedexecnarr) -- node {abstract types and API calls} (execnarr2proverif);
	\path [line] (execnarr2proverif) --  node {applied pi-calculus verification} (proverif);  
    \path [line,dashed,pos=0.4] (proverif) -| node {unsafe} (revise);
    \path [line,dashed] (revise) -- ++(0,2.5) node[yshift=0.5cm] {AnBx Protocol Modelling};
	\path [line] (proverif) -- node {safe} (javacodegen);
	\path [line] (javacodegen) -- node[xshift=-4.5cm] {concrete types and API calls} (javacode);
    \path [line] (javatemplates) -- (javacodegen);
	\path [line] (typedexecnarr) -- ++(-2.6,0) -- ++(0,-2) -- ++(0,-1) --(javacodegen);

\end{tikzpicture} 
}
\par\end{centering}
\caption{\label{fig:Code-Generation}\anbx{} Compiler back-end (adapted from \cite{anbx2015})}

\end{figure}

We briefly present the \anbx{} Compiler and Code Generator by illustrating the main steps involved in the automatic Java code generation for security protocols. A more detailed description can be found in \cite{anbx2015}.

The \anbx{} input protocol undergoes lexing and parsing before being compiled into \anb{}, a format that enables automated verification with the OFMC model checker. If the verification process is successful, the \anb{} specification can be translated into an \emph{executable narration} (\execnarr{}), which operationally encodes the sequence of actions that agents are expected to perform during the protocol. This step specifically involves computing checks on receptions, actions that agents must execute on incoming messages to ensure that the protocol adheres to the specification. For instance, this includes the verification of digital signatures, decryption of incoming messages, and equality tests to confirm that incoming messages match prior knowledge.

Following this, the sequence of actions can be reordered into an \emph{optimised executable narration} (\optexecnarr{}) using optimisation techniques such as common sub-expression elimination (CSE). This optimisation minimises the number of cryptographic operations and reduces overall execution time.

The Java code generation requires an additional step, as illustrated in Figure \ref{fig:Code-Generation}. We model the protocol logic using a language-independent intermediate format called \jprotocol{}. This format serves as a typed representation of the \optexecnarr{}, allowing for parametrised translation and simplifying code emission in the target programming language.

The type system of \jprotocol{} infers the types of expressions and variables, ensuring that the generated code is well-typed. To that end, the \anbx{} compiler utilises, along the entire compilation chain, the function signature typing and other language extension information peculiar to \anbx{}, which are not available in \anb{}. It also provides the advantage of runtime checks to verify whether the structure of incoming messages matches the expected format as specified by the protocol. 

Java code emission is accomplished by instantiating protocol templates (referred to as \emph{application logic}), which serve as the skeleton of the application. It is important to note that only at this final stage do language-specific features and API calls bind to the protocol logic.

The verification of the protocol logic can be conducted using ProVerif, which translates from \jprotocol{} to the applied-pi calculus \cite{blanchet2001efficient}.

In summary, ProVerif verifies the symbolic implementation from which the Java code is emitted, while OFMC verifies the abstract model created by the designer.

\subsection{Security protocol verifiers}

Utilising several verification tools is often beneficial for cross-checking the validity of security properties under different assumptions. Galeotti \textit{et al.} \cite{DBLP:journals/corr/GaleottiFMFZ14} demonstrate an integration of multiple tools into a single framework that addresses the limitations of individual tools. Some differences between verification tools are detailed in \cite{DBLP:conf/sp/BarbosaBBBCLP21}, with updates reflecting the most recent developments in ProVerif.

\textbf{OFMC} \cite{basin2005ofmc} utilises the AVISPA Intermediate Format (IF) \cite{avispa-manual} as its native input language. IF enables the description of security protocols as infinite-state transition systems using set rewriting. Additionally, OFMC supports the \anb{} language. The tool performs both protocol falsification and bounded session verification by exploring the transition system in a demand-driven manner. The major techniques employed by OFMC include the \textit{lazy intruder}, a symbolic representation of the intruder, and \textit{constraint differentiation}, a search-reduction technique that combines the lazy intruder with concepts from partial-order reduction. This integration effectively reduces the search space without excluding potential attacks or introducing new ones.

\textbf{ProVerif} \cite{blanchet2001efficient} is an automated verifier for cryptographic protocols that models both the protocol and the attacker using the Dolev-Yao \cite{dolev83ieee} symbolic approach. This approach represents data and ideal cryptographic operations symbolically, assuming the attacker has complete control over public communication channels. Unlike model checkers, ProVerif can model and analyse an unbounded number of parallel sessions of the protocol. However, similar to model checkers, ProVerif can reconstruct a possible attack trace when it detects a violation of the intended security properties. While ProVerif may report false attacks, if a security property is reported as satisfied, it is guaranteed to hold in all cases. Therefore, careful analysis of the results is crucial when attacks are reported.\\

In our context, it is important to note that OFMC is sound and complete, which, in practical terms, implies that if there is any attack on the protocol, the tool will detect it, and no false attacks can be reported. In contrast, ProVerif is only sound, meaning that potentially false attacks can be reported, for example, due to over-approximation. While it is remarkable that this tool can verify protocols for an unbounded number of sessions, non-termination or undecidability may occur in some cases. Unlike OFMC, ProVerif can check diff-equivalence properties \cite{DBLP:journals/jlp/DelauneH17}, determining if protocols share the same structure while differing only in the messages they exchange. Conversely, OFMC accepts a simpler input language (i.e., \anb{}) and provides an attack trace that is easier to interpret. Furthermore, ProVerif may flag certain goals as satisfied even when the associated protocol steps are unreachable due to errors in the encoding of the model, which can lead to confusion regarding the protocol's security.

\section{Methodology}
\label{sec:methodology}

\begin{figure}[!h]
\begin{centering}
\resizebox{80mm}{!}{

\tikzstyle{decision} = [diamond, draw, fill=blue!10,text width=4.75em, text badly centered, node distance=1.5cm, inner sep=0pt,font=\sffamily]
\tikzstyle{block} = [rectangle, draw, fill=blue!10,text width=10em, text centered, rounded corners, minimum height=3em, minimum width=3cm,font=\sffamily]
\tikzstyle{line} = [draw, -latex',font=\sffamily]
\tikzstyle{cloud} = [draw, ellipse,fill=red!20, node distance=1.5cm, minimum height=2em, minimum width=4cm, font=\sffamily]

\begin{tikzpicture}[node distance = 1.5cm, auto]
        \node [cloud] (mentalmodel) {Mental modelling};

    \node [block, below of=mentalmodel] (requirements) {Informal Specification and Requirements};
    \node [cloud, below of=requirements] (modelling) {Formal Modelling};
    \node [block, below of=modelling] (anbx) {AnBx Protocol Model};
	\node [cloud, below of=anbx] (anbxc1) {AnBxC Compilation};
	\node [block, below left of=anbxc1, node distance=3cm] (anb) {AnB Protocol Model};
 	\node [cloud, below of=anb] (ofmc) {OFMC};
 	\node [block, below right of=anbxc1, node distance=3cm] (proverifmodel) {ProVerif\\ Protocol Model};
	\node [cloud, below of=proverifmodel] (proverif) {ProVerif};

	\node [cloud, below of=anbxc1,node distance = 5cm] (anbxc2) {AnBxC Back-End};
    \node [block, below left of=anbxc2,,node distance = 3cm] (javacode) {Java Code};

      \node [block, below right of=anbxc2,node distance = 3cm] (javadocker) {Dockerised Java};
    
    \node [cloud, below of=anbxc2,node distance = 4cm] (exec) {Implementation execution};
        \path [line,dashed] (mentalmodel) -- (requirements);

    \path [line,dashed] (requirements) -- (modelling);
	\path [>=stealth,line,dashed,<->]  (modelling) -- (anbx);
	\path [line] (anbx) -- (anbxc1);
	\path [line] (anbxc1) -- (anb);
 	\path [line] (anbxc1) -- (proverifmodel);
 	\path [line] (proverifmodel) -- (proverif);

    \path [line] (anb) -- (ofmc);

    \path [line] (ofmc) -- node[pos=0.5,left] {safe} (anbxc2);
        \path [line] (proverif) -- node[pos=0.5,right] {safe} (anbxc2);

	\path [line,dashed,pos=0.03] (proverif.east) -- ($(proverif.east)+(1,0)$)  |- node {unsafe} (mentalmodel);
 	\path [line,dashed,pos=0.03] (ofmc.west) -- ($(ofmc.west)+(-1,0)$) |- node[right] {unsafe} (mentalmodel);

	\path [>=stealth,line,dashed,<->] (anb.west)  -- ($(anb.west)+(-0.5,0)$)  |- (modelling.west);

	\path [>=stealth,line,dashed,<->] (proverifmodel)  -- ($(proverifmodel.east)+(0.5,0)$)  |- (modelling.east);
 
   \path [line] (anbxc1) --  (anbxc2);

    \path[line] (anbxc2) -- (javacode);

    \path[line] (anbxc2) -- (javadocker);
    \path[line] (javadocker) -- (exec);
    \path[line] (javacode) -- (exec);

\end{tikzpicture} 

}
\par\end{centering}
\caption{\label{fig:idemethodology}Model Driven Development with the \anbx{} IDE
(\protect\tikz[baseline]{\protect\draw[dashed] (0,.8ex)--++(1,0);}
manual\quad{}\protect\tikz[baseline]{\protect\draw[] (0,.8ex)--++(1,0);}
automatic)}
\end{figure}
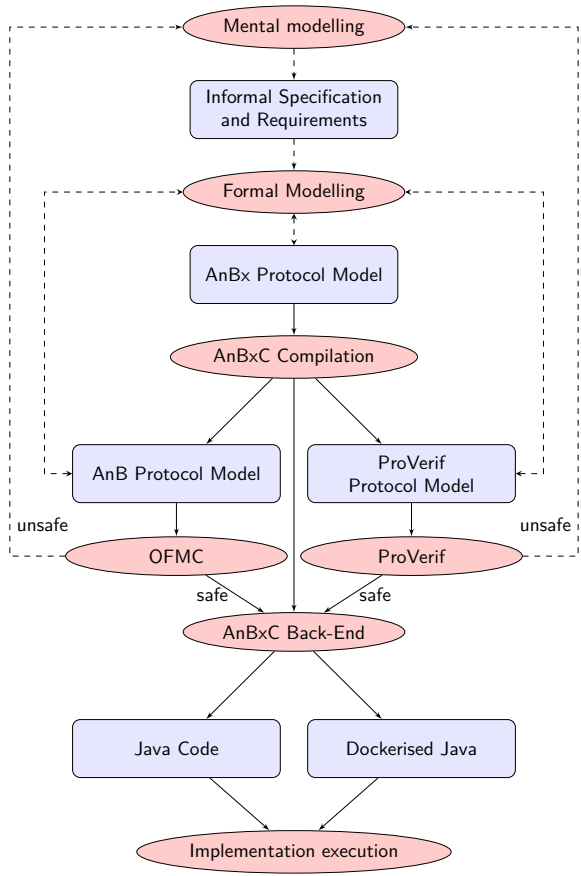

The methodology that our IDE intends to support follows the classical Model-Driven Development (MDD) philosophy. 

The process involves specifying an abstract model of a security protocol, reasoning about its correctness and soundness with the aid of verification tools, and ultimately generating a concrete executable implementation. The MDD methodology, applied in the context of our IDE, is detailed in Figure \ref{fig:idemethodology}, and our contribution focuses on reducing friction between each step of the process to make it user-friendly. Achieving this requires automating the design and development workflow, as well as supporting users with early warnings, error messages, and suggestions to identify, correct, and prevent mistakes.

While every step in this process would typically have to be performed manually, or at best using scripts, the IDE can automatically launch a chain of compilation, verification, implementation, and execution steps with just a couple of keystrokes. This makes the process flexible and approachable for beginners. 

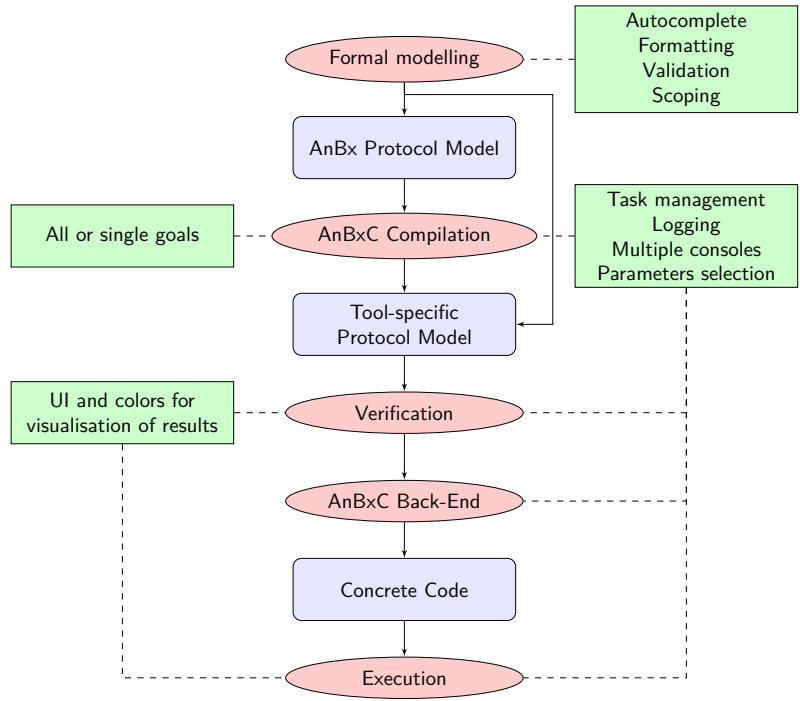
\begin{figure}[!h]
\begin{centering}
\resizebox{110mm}{!}{

\tikzstyle{decision} = [diamond, draw, fill=blue!10,text width=4.75em, text badly centered, node distance=1.5cm, inner sep=0pt,font=\sffamily]
\tikzstyle{block} = [rectangle, draw, fill=blue!10,text width=10em, text centered, rounded corners, minimum height=3em, minimum width=3cm,font=\sffamily]
\tikzstyle{line} = [draw, -latex',font=\sffamily]
\tikzstyle{cloud} = [draw, ellipse,fill=red!20, node distance=1.5cm, minimum height=2em, minimum width=4cm, font=\sffamily]
\tikzstyle{idecontrol} = [draw, rectangle,fill=green!20, text width=10em, text centered, minimum height=3em, minimum width=2cm,font=\sffamily]

\begin{tikzpicture}[node distance = 1.5cm, auto]

    	\node [cloud] (modelling) {Formal modelling};

    \node [idecontrol,right of=modelling,node distance=4.75cm] (idemessages) {Autocomplete\\Formatting\\Validation\\Scoping};

    \node [block, below of=modelling] (anbx) {AnBx Protocol Model};
	\node [cloud, below of=anbx] (anbxc1) {AnBxC Compilation};

 	\node [idecontrol,right of=anbxc1,node distance=4.75cm] (taskmanagement) {Task management\\Logging\\Multiple consoles\\Parameters selection};
  
 \node [idecontrol,left of=anbxc1,node distance=4.75cm] (singlegoals) {All or single goals};

	\node [block, below of=anbxc1] (toolspecific) {Tool-specific\\Protocol Model};

 	\node [cloud, below of=toolspecific] (verification) {Verification};
    \node [idecontrol,left of=verification,node distance=4.75cm] (visualisation) {UI and colors for visualisation of results};

     	\node [cloud, below of=verification] (implementation) {AnBxC Back-End};

	\node [block, below of=implementation] (concretecode) {Concrete Code};

 	\node [cloud, below of=concretecode] (execution) {Execution};

   	\draw [dashed] (idemessages) -- (modelling);
	\path [line] (modelling) -- (anbx);
	\path [line] (modelling) |- ($(modelling)+(2.5,-0.6)$) |- (toolspecific);
1
	\path [line] (anbx) -- (anbxc1);
	\path [line] (anbxc1) -- (toolspecific);
	\draw [dashed] (taskmanagement) -- (anbxc1);
  	\draw [dashed] (singlegoals) -- (anbxc1);

  	\draw [dashed] (taskmanagement) |- (verification);
  	\draw [dashed] (taskmanagement) |- (implementation);
     	\draw [dashed] (visualisation) |- (execution);

  	\draw [dashed] (taskmanagement) |- (execution);

    \path [line] (toolspecific) -- (verification);

    \draw [dashed] (visualisation) -- (verification);

    \path [line] (verification) -- (implementation);
    \path [line] (implementation) -- (concretecode);
    \path [line] (concretecode) -- (execution);

\end{tikzpicture} 

}
\par\end{centering}
\caption{\label{fig:idefeatures}Control and visualisation over the modelling and verification process with the \anbx{} IDE (\protect\tikz[baseline]{\protect\draw[dashed] (0,.8ex)--++(1,0);}
IDE features\quad{}\protect\tikz[baseline]{\protect\draw[] (0,.8ex)--++(1,0);}
workflow)}
\end{figure}

The integration advantages and features offered by the IDE are summarised in Figure \ref{fig:idefeatures}. The editor enables the modelling of protocols with features like autocomplete, formatting, validation, and scoping. Then compilation, verification, and execution procedures are supported with task management, logging, and console output separation. After verification or execution, the IDE provides clear visualisation, with console output colouring and a dedicated Eclipse view window.

\section{\anbx{} IDE Features and Components}
\label{sec:idecontribution}

In this section, we detail how we cover the entire process of writing, verifying, generating, and running code within the same environment, with supporting features and components. Our approach integrates language and interface for a complete Model-Driven Development (MDD) workflow that remains user-friendly.

\subsection{Editing  Security Protocol models in different specification languages}
\label{subsec:ideediting}

Several grammars are defined in the \anbx{} IDE to encode security protocols in different specification languages: Alice \& Bob notation (\anb{} and \anbx{}), applied-pi calculus (PV) and AVISPA/OFMC Intermediate Format (IF).

As we advocate for a Model-Driven Development (MDD) approach, the \anb{} notation and its extension \anbx{} are fully supported, along with common IDE features that we detail later in this section. 

This support extends not only to \anb{}-like notations but also to the additional languages of the verification tools, namely applied-pi for ProVerif, and specifications in OFMC Intermediate Format (IF) and Theory (thy) formats. This enables users to comfortably edit more fine-grained specifications than allowed in \anb{} style, thereby becoming better acquainted with the tools' inner workings.

 \subsubsection{Getting started with wizards}\label{subsec:ide-wizard}   

To get started before writing a specification, the user can utilise appropriate wizards. The IDE features wizards for \anbx{} and ProVerif projects, which can be accessed with just a few clicks through the \texttt{File$\rightarrow$New} menu commands. A stub is generated, akin to the "Hello world" code in other programming languages. Similar wizards also allow for the creation of single \anbx{} or ProVerif files without generating an entire project structure.

 \subsubsection{Syntax highlighting and outline}\label{subsec:ide-syntax} 

For easier visualisation, syntax highlighting is applied to the specifications. This feature aids in locating sections and keywords, and it can be customised thanks to XText's internal \emph{DefaultSemanticHighlightingCalculator}, which can be extended at will. Specifically, we have chosen a red colour for section names, while keywords receive the default purple highlighting. All supported languages benefit from this highlighting.

Additionally, for all languages in the \anbx{} IDE, we provide an outline of the specification for visualisation at a glance. This outline summarises the structure and core aspects of the currently edited file, in a dedicated collapsible tree view within Eclipse.

 \subsubsection{Formatting} \label{subsec:ide-formatting}

To maintain a clear structure and continue with the visualisation features, the IDE can auto-format \anb{}/\anbx{} files. Spaces between tokens, line breaks, and indentations are enforced, ensuring that the readability benefits of using a high-level notation are not diminished after long editing sessions. This formatting can be triggered through the defined Eclipse keyboard shortcuts, resulting in a structure similar to that shown in Figures \ref{fig:AnBx-Protocol-Example} and \ref{fig:AnB-Protocol-Example}.

 \subsubsection{Autocomplete and scoping}\label{subsec:ide-autocomplete}

When writing any specification, autocomplete is one of the most essential features. In our supported languages, declared entities and keywords are suggested to the developer, along with other symbols that conform to the language grammar at a specific point.

To determine what can legitimately be used at a given time in the specification, a notion of scoping needs to be introduced in most languages. For example, in \anbx{}, one can use macros, and their parameters must be local to the macro. Similar requirements exist in ProVerif, albeit in a more complex manner, where bound variables and multiple processes can be used. Scoping for both languages is fully supported in our plug-in.

\begin{figure}[!h]
\begin{centering}
 \includegraphics[scale=0.8]{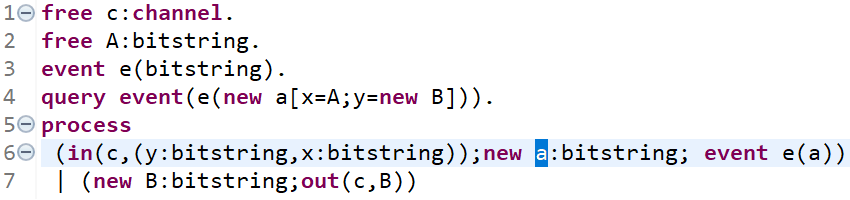}
\par\end{centering}
\caption{\label{fig:pvscoping}Tracking declarations with ProVerif scoping}
\end{figure}

An appropriate use of scoping is particularly beneficial for specifications that might be confusing at first glance. We take an example from the ProVerif manual \cite{blanchetproverifmanual} in Figure \ref{fig:pvscoping} to illustrate our point.

The query in line 4 refers to previously declared names such as event \texttt{e} and the free name \texttt{A}. However, \texttt{new a}, \texttt{x}, \texttt{y}, and \texttt{new B} are forward references. Pointing to one of them would take us to their declaration locations, specifically line 6 for \texttt{a} and line 7 for \texttt{B}. ProVerif also employs bindings for variables that exist within a certain scope. In this context, the square brackets for \texttt{new a} constrain the values of \texttt{x} and \texttt{y} to equal \texttt{A} and \texttt{new B} respectively, within the scope accessible by \texttt{new a}. Without a solid understanding of ProVerif applied-pi, this specification could appear obscure. Therefore, declaration tracking through scoping serves to clarify these relationships.

\subsubsection{Validation: type, arity, and semantics checking with quickfixes}\label{subsec:ide-validation}

As with many other standard IDEs, we provide type and arity checking. Functions and other language features can exhibit a signature, which must be enforced without the need to run the verification tools. In addition to what the user explicitly defines, a set of predefined symbols and signatures is taken into account. For example, \anb{} protocols rely on primitives such as \texttt{inv}, \texttt{exp}, and \texttt{xor} to denote a private key, exponentiation, and exclusive-or. Meanwhile, \anbx{} protocols also rely on public key functions \texttt{pk} and \texttt{sk}, as well as a symbolic \texttt{hash} function. ProVerif includes predefined types such as \texttt{bitstring}, \texttt{bool}, and \texttt{nat}. These signatures and symbols are hard-coded as libraries that a specification can use as if they were defined in the current file.\\

\begin{figure}[!h]
\begin{centering}
\includegraphics[scale=0.45]{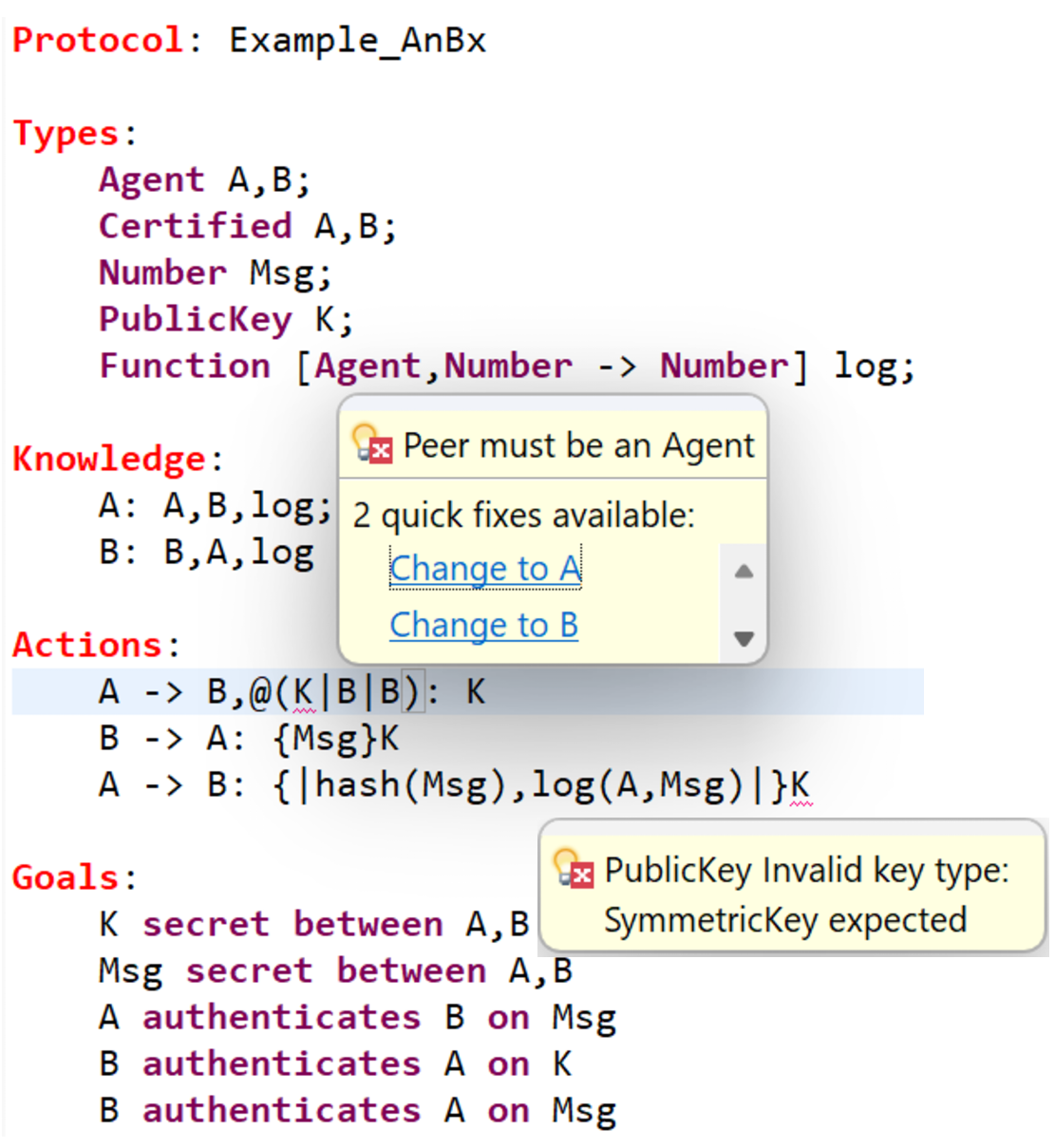}
\par\end{centering}
\caption{\label{fig:Quick-Fix}Error message and quick fix}
\end{figure}

The IDE enforces many sanity checks and signals any violations to the user through the XText \emph{Validator} component. Moreover, the validation feature notably handles arity of function calls and type-checking. Taking the protocol defined in Figure \ref{fig:Quick-Fix} as an example, a function signature like \textbf{$\mathtt{log:Agent,Number \rightarrow Number}$} will be compared to its calls, accepting only \texttt{Agent} or a function returning an agent as the first parameter. Similarly, a call to \texttt{log} cannot be used as a parameter unless the called function's signature expects a \texttt{Number}.

The protocol in Figure \ref{fig:Quick-Fix} contains two errors. The first error arises because a channel mode only accepts parameters of type \texttt{Agent}. The type checker returns the cause of the error and provides two suggestions on how to resolve it. The second error occurs due to the use of illegal cryptographic notations. In fact, a challenge for a developer is determining which encryption to use in specific situations. The validation system guarantees the correctness of types according to the cryptographic specifications. The user is attempting to use a public key in a symmetric cipher scheme. To fix this error, the user has two possibilities: changing the type of \texttt{K} to \texttt{SymmetricKey} or changing the encryption scheme from symmetric to asymmetric. However, the protocol is safe only in the first case, while the second case introduces a potential attack. This can be determined by verifying the protocol.\\

These examples demonstrate how the IDE can assist users with limited knowledge of cryptography in avoiding design mistakes. The use of the most basic primitives, such as \texttt{pk}, might seem clear to someone familiar with the language. However, even understanding that this symbol represents the key of a particular agent, and that only this agent is entitled to it, may not be evident when described as a public value. During our practical activities with students, some attempted to model a shared symmetric key as \texttt{pk(A,B)}, confusing symmetric and asymmetric schemes, and still under the impression that one key must be used and shared by both the sender and the receiver with public-key cryptography. Instead \texttt{pk} is a function that is used for asymmetric encryption and expects a single parameter of type \texttt{Agent}. The feedback in the editor is immediate, eliminating the need to run the compiler and track line and column numbers, resulting in a short-lived misconception.

\begin{figure}[h]
\begin{centering}
\includegraphics[scale=0.5]{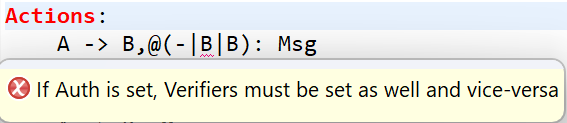}
\par\end{centering}
\caption{\label{fig:Auth-and-Verifiers}Auth and Verifiers checking}
\end{figure}

The editor can as well help to avoid possible mistakes with the \anbx{} channel's syntax, which consists of a triple (\emph{auth}$\mid$\emph{vers}$\mid$\emph{dest}). For example, the \emph{auth} parameter (the agent authenticating a message, e.g., with a signature) and the \emph{vers} parameter (a set of agents that the \emph{auth} intends to enable to verify the authenticity of the message) can only occur together in a single channel mode but not individually (see Figure \ref{fig:Auth-and-Verifiers}).

\subsection{Running the verification tasks and generating code}\label{subsec:ide-running}

Once the user has specified the protocol according to their vision, benefiting from editor feedback, a combination of UI components and workflow helpers can be employed. These features provide flexible and convenient control over the upcoming compilation and verification tasks.

\subsubsection{Configuration dialogs with options and help} \label{subsec:ide-configuration-dialogs}

The integration of design and verification tools begins with dedicated UI windows that allow users to set command-line parameters and execute common tasks with just a few clicks.

For example, these dialogs feature dropdown lists, spinners, and checkboxes for the most common options, and facilitate interaction across tools. An \textit{AnBxC additional args} textbox is provided for each tool, enabling users to specify any command-line parameters without cluttering the  interface.

\begin{figure}[h]
\begin{centering}
\includegraphics[scale=0.75]{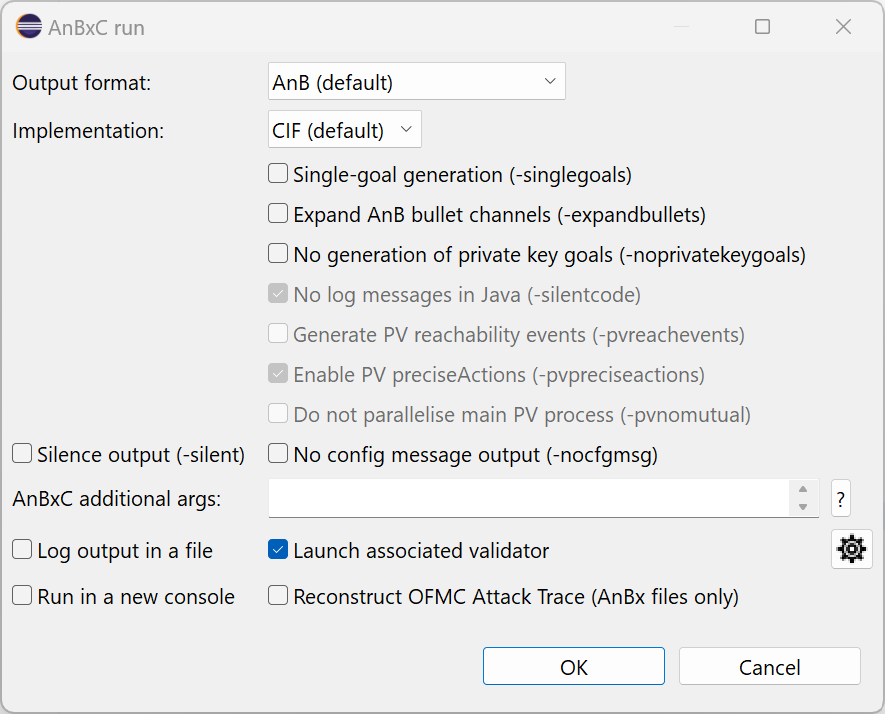}
\par\end{centering}
\caption{\label{fig:anbxcdialog}AnBxC dialog}
\end{figure}

Next to this textbox, we include a help button that opens the manual for the relevant tool. The window that appears contains a link to the complete official documentation, along with a scrollable, selectable text area detailing the options, presented as the tool would display them in a terminal when requesting help. An example of tool dialog is given in Figure \ref{fig:anbxcdialog}, featuring the one for AnBxC.

The \textit{Launch associated validator} option triggers automatic verification with OFMC or ProVerif, depending on the chosen output format. It is important to note that enabling a real-time modelling and verification feedback cycle is crucial for increasing productivity when dealing with complex protocols. Adjacent to this option, a settings button opens the dialog for the associated verification tool, namely OFMC or ProVerif, allowing users to configure verification options for the next run without unnecessary back-and-forth.

As the \anbx{} Compiler enables users to generate a Java implementation for the specified protocol, the same option triggers the building and execution of the generated code. This process is orchestrated by an Ant file, a standard build file for the Eclipse platform and other IDEs. Alternatively, the associated Java project can be opened in Eclipse, where users can manually run its Ant build file. Observing the execution trace at a concrete level after working in \anbx{} is essential, as it allows users to understand how abstract concepts are translated into the real world. This understanding is crucial for grasping the nuances of cryptographic operations and their practical applications.

\subsubsection{Supporting users in their workflow} \label{subsec:ide-supporting-users}

Opting for an integrated development environment is beneficial for automating intermediate steps and saving considerable time, avoiding unnecessary back-and-forth.

Typically, the running time of OFMC increases drastically with the number of sessions. For fast prototyping, it is important to first verify that our specification satisfies the security goals in a single session and then verify it again with multiple concurrent sessions. The IDE features an option for this in the OFMC dialog. The \textit{Run with one session first} option attempts verification according to the options specified in the dialog, with the session number set to 1. If an attack is found, the process stops here, and verification with concurrent sessions only occurs if no attack is identified.

Intermediate steps with OFMC may also arise when we want to generate an IF representation of the \anb{} protocol and then verify it. By using a simple \textit{Run on an on-the-fly generated IF file} option, the generation and verification commands are executed sequentially.

Another important aspect of OFMC verification is the use of equational theories. These are specified in .thy files, but we can automatically generate these files using the \anbx{} Compiler along with appropriate \anbx{} notations. From an \anbx{} file, the compiler can output a theory file that encodes the relations in the \texttt{Equations} section, allowing non-experts to reason with custom functions while saving considerable time for more advanced users. Understanding the internals of OFMC and how to encode equational theories for this particular tool is not necessary. Users can then concentrate on the core aspects of the logic and avoid wasting time that could be better spent on understanding the reasoning behind such relations. Appropriate output formats and options are available in the \anbx{} Compiler and OFMC dialogs to generate and include these theory files.

\subsubsection{Verifying single or multiple security goals in parallel} \label{subsec:ide-parallel-goals}
\label{sec:idesinglegoals}

The \textit{Single-goal generation} option (Figure \ref{fig:anbxcdialog}) generates files in the target format as if we had a separate copy of the protocol for each goal. The verification tools can then run on every generated protocol and test a single property. Compared to checking all goals within a single process, our experience shows that the overhead of spawning multiple processes is reasonable and well worth it when aiming to obtain precise information on each goal as quickly as possible. Moreover, this approach takes advantage of multicore CPUs, offering significant time savings through parallelisation.

With OFMC, it is particularly useful to quickly identify which goals have failed, as it typically reports the first attack trace it finds. Although the output generated by OFMC clearly indicates the type of goal that has been violated (e.g., authentication, secrecy), it can be much more challenging to determine the exact vulnerable goal in cases where multiple goals of the same category fail.

This solution with single goals helps to maintain a readable and concise verification output. Moreover, OFMC operates with a bounded number of sessions. If we obtain a result for one session and subsequently wish to check whether all goals can be satisfied with multiple sessions, this parallel approach can yield results considerably faster than waiting for all the goals of a complex protocol to be checked sequentially. We observe a particular parallelisation advantage in this regard in the case of ProVerif. 

It should be noted that in ProVerif, some goals can be deemed satisfied when the associated protocol steps are unreachable, as the goals are specified as implication queries. Indeed, not reaching the point where we could have \( A \) makes the expression \( A \implies B \) evaluate to true. This can lead to confusion for the user, who might write a specification considered correct while it is not executable by design. Ensuring that all agents can properly terminate their respective processes requires the presence of events in the ProVerif specification. The \anbx{} dialog offers an option to do this automatically when generating a ProVerif file. Consequently, a user would be more likely to recognise the benefits of these events instead of being potentially misled by ProVerif results.\\

\begin{table}[t]
\makebox[\textwidth][c]{
\small
\begin{threeparttable}
\begin{tabular}{ lcS[table-format=3.2]S[table-format=3.2]S[table-format=4.2]S[table-format=1.3]S[table-format=1.3]S[table-format=3.2] }
\toprule
 \multirow{2}{*}{\textbf{Protocol}} & \textbf{Goals/}  & \multicolumn{3}{c}{\textbf{OFMC (2 Sessions)}} & \multicolumn{3}{c}{\textbf{ProVerif (unlimited S.)}} \\
  \cmidrule{3-8}
 & \textbf{Attacks} & \boldmath{$all$} &  \boldmath{$sgl$} &  \boldmath{$\Delta$ \%} &  \boldmath{$all$} &  \boldmath{$sgl$} &  \boldmath{$\Delta$ \%} \\ 
\midrule

 Carlsen \cite{DBLP:journals/sigops/Carlsen94,ofmcsite} & 3/0 & 5.64 & 6.07 & 7.62 & 0.090 & 0.032 & -64.44 \\ 
\midrule

   H530  \cite{h530,ofmcsite} & 2/2 & 16.05 & 19.14 & 19.25 & 0.257 & 0.081 & -68.48 \\ 
\midrule
  
  IKEv2DS \cite{kaufman2005internet,ofmcsite} & 4/0 & 32.86 & 34.78 & 5.84 & 2.774 & 0.980 & -64.67 \\
\midrule
  ISO5Pass \cite{isosymkey,ofmcsite} & 3/2  & 1.22  & 23.25  & 1805.74 & 0.337 & 0.093 & -72.40 \\
\midrule
  ISOCCF3PassMutual \cite{isoccf,ofmcsite} & 2/0 & 3.47 & 4.19 & 20.75 & 0.042 & 0.021 & -50.00 \\
\midrule

 NSL \cite{DBLP:journals/stp/Lowe96,ofmcsite} & 2/0 & 0.78 & 0.81 & 3.85 & 0.062 & 0.028 & -54.84 \\ 
\midrule
 
 NSPK \cite{DBLP:journals/cacm/NeedhamS78,ofmcsite} & 4/4 & 0.32 & 0.88 & 175.00 & 0.077 & 0.027 & -64.94  \\ 
\midrule
 
  Otway-Rees \cite{DBLP:journals/sigops/OtwayR87,ofmcsite} & 3/0 & 16.78 & 17.38 & 3.58 & 0.116 & 0.038 & -64.24  \\ 
\midrule
  TLS \cite{DBLP:journals/tissec/Paulson99,ofmcsite} & 3/0 & 70.24 & 68.69 & -2.21 & 0.504 & 0.068 & -85.51 \\ 
\midrule
 Woo-Lam 92 \cite{woo1992authentication,ofmcsite} & 3/1 & 266.57 & 325.21 & 22.00 & 8.329 & 2.401 & -71.17 \\
\midrule
  Yahalom \cite{paulson2000yahalom,ofmcsite} & 6/1 & 2.38 & 4.28 & 79.83 & 0.131 & 0.030 & -77.10 \\
\midrule
\end{tabular}
 \begin{tablenotes}
        \item[] $all$ -- All-goals verification time  \item[] $sgl$ -- Parallel single-goals verification time     
        \item[] $\Delta$ \% -- Difference in verification time (\%): $100*(sgl-all)/all$
    \end{tablenotes}
    \end{threeparttable}
}
 \caption{Comparison of verification times (for typed models) in seconds and variation between single-goal and all-goal verification for OFMC and ProVerif.}
 \label{tab:singegoalsbenckmarktable}

\end{table}

We benchmark the overhead of the single goals feature, and we report results in Table \ref{tab:singegoalsbenckmarktable}. The benchmarking was conducted on an Intel(R) Xeon(R) W-2145 CPU running at 3.70 GHz, equipped with 128 GB of RAM and operating on Windows 10 64-bit. We tested OFMC 2022 with two sessions, and ProVerif 2.05 for an unbounded number of sessions. We used half of the CPU cores to ensure smooth task progression and maintain a good balance in overall machine performance, and at least 6 cores were available to perform the single goal verification tasks in parallel. Tasks were repeated at least 20 times, and then we computed the average verification times.

With these benchmarks, one may immediately note that single-goal parallel verification is always advantageous for ProVerif's performance. Moreover, this approach may help to detect reachability issues more easily. In all tested cases, the verification time is reduced by at least 50\%.

As expected, given the verification technique used by OFMC, there is an overhead on almost all cases. While the most evident convenience lies in clearly indicating in the results which goal is failing, the advantage of parallel single-goal verification is that all goals can be checked instead of halting at the first failing one. In percentage terms, the large discrepancies between parallel single and all-goal verification times are usually due to OFMC stopping at the first failing goal in the all-goals verification scenario. 

As a reference, performing single-goal verification manually would introduce significant overhead, requiring the user to launch each task individually. In the worst-case scenario, tasks would run sequentially; at best, extra time would be needed to enter each command and execute them as background processes. This manual approach, when employing the command line, also results in reduced visualisation of verification results and less control over running tasks, as it relies on the operating system’s task manager rather than the IDE’s dedicated task manager.

\subsubsection{Java Code Generation and Run of Dockerised applications} \label{subsec:ide-java}

The \anbx{} Compiler can generate Java code from an \anbx{} specification, and the IDE can coordinate the execution of the distributed application using an automatically generated Ant file. In the most basic setting, the application will run using the localhost interface; however, recent developments in the compiler have extended its features to allow for the generation of Dockerised distributed Java code. Docker is a platform that enables developers to automate the deployment of applications inside lightweight, portable containers. Each agent's code can now run on different machines, which allows for greater flexibility and scalability. Furthermore, the IDE can execute this Dockerised Java in just a few steps, streamlining the process for users.

When the export to Java (Docker) option is selected, the call to the compiler is followed by several Docker commands that manage the lifecycle of the application containers. These commands are as follows:

\begin{itemize}
	\item \texttt{docker compose \emph{lastBuildFile} down}: This command stops and removes the containers defined in the specified Docker Compose file, which represents the last built project. This ensures that any previously running instances of the application are halted before starting a new one.
	\item \texttt{docker container prune}: This command removes all stopped containers, helping to free up system resources and reduce clutter in the Docker environment.
	\item \texttt{docker network prune}: This command cleans up unused networks, ensuring that only necessary connections are maintained within the Docker ecosystem.
	\item \texttt{docker compose \emph{currentBuildFile} up}: This command starts the containers defined in the current Docker Compose file, effectively launching the application in its new environment.
\end{itemize}

Here, \emph{lastBuildFile} and \emph{currentBuildFile} refer to YAML Docker Compose files that are tied to the last and currently built Docker projects, respectively.

Essentially, this sequence of commands stops previous containers, deletes unused containers and networks from the disk, and runs the container of interest. This approach ensures that no other \anbx{}-related container interferes with the workflow.

\subsubsection{Console output and logging} \label{subsec:consoleoutput}

At the bottom of any tool dialog (e.g., Figure \ref{fig:anbxcdialog}), users can choose to run the next process in a new console. This feature helps to separate the outputs of different tasks, providing a clearer view of the unfolding results. This option leverages standard Eclipse consoles, which can be selected like any other through the console view's toolbar menu. The consoles are numbered and have their own stream processing, ensuring that the output from one console does not interfere with that of another.

To interpret console output at a glance, we colour it according to pre-defined regular expressions. For example, a detected attack is coloured red, while the absence of an attack on a given protocol or goal is represented in green, and undecidability is indicated with an orange output. The policy applies to all tools printing to an \anbx{} IDE console and can be deactivated in the configuration window.

To facilitate output analysis, we can go beyond merely compartmentalising it into several consoles by saving it in dedicated log files with another option. These log files are named after the protocol, a runtime timestamp, and export options when relevant. Each log file is located in a folder named after the tool that generated the output, making retrieval straightforward. Users can then launch an entire suite of tasks and later perform statistical analyses on the logs.

From a research perspective, these two options are invaluable when comparing multiple prototypes for runtime, correctness, etc. They also help in building a database of case studies for protocols verified as secure or subject to interesting attacks.

  \subsubsection{Displaying verification results in an Eclipse view}
\label{subsec:ideverifresultsview}
With multiple tasks running and reaching conclusions concurrently, we help the user by providing a clear and concise summary of these results. To achieve this, we use a dedicated Eclipse view of the verification results, which details, in a collapsible tree view, for each protocol by file name, which goals succeeded in \textit{green}, failed in \textit{red}, or are deemed undecidable in \textit{orange}.

\begin{figure}[!h]
\begin{centering}
\includegraphics[scale=0.9]{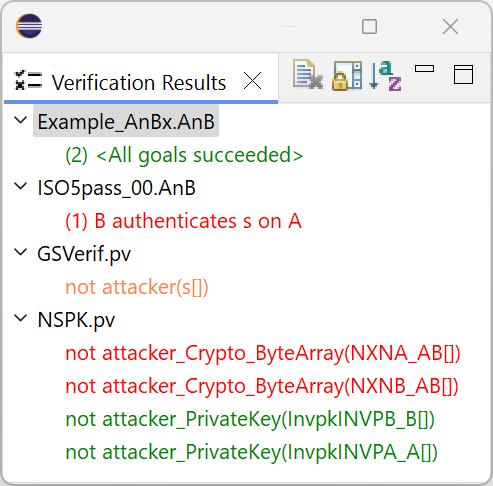}
\par\end{centering}
\caption{\label{fig:ideverifresults}View of verification results}
\end{figure}

The targeted goals are colour-coded, session-annotated, and sorted with the failing ones first. The user also has the option to sort the list of protocols alphabetically, enable scroll lock (similarly to a console), or clear the list. An example is shown in Figure \ref{fig:ideverifresults}.

With this integrated view, there is no need to look in detail at the console output or scroll through long blocks of information to understand the current status. For ProVerif, the view displays precise extracts of the output, making it easily searchable when more detail is required.

 \subsubsection{Scheduling with priorities and Task manager}
 \label{subsec:taskmanager}
As users launch many tasks, the order of execution needs to be considered for productivity. The tasks are organised in a queue, and we implement a priority policy for the waiting tasks.

A user may want to experiment with some light verification or generation while a more time-consuming task is running. Our approach is to move lighter tasks up the queue, thereby avoiding the need to wait for the heavier ones to complete. We have implemented a priority queue as detailed in Algorithm \ref{alg:task_scheduling}: the compiling tasks are executed first, followed by OFMC verification with one session, then ProVerif, and finally OFMC with multiple sessions. Tasks with the same priority are executed in a \emph{First-In, First-Out} (FIFO) fashion.

\begin{algorithm}
	\caption{Task Scheduling with Priority Queue}
	\label{alg:task_scheduling}
	Initialise an empty priority queue $Q$\;
	Define priority levels\;
	$P_1 \gets$ Highest priority: Compiling tasks\;
	$P_2 \gets$ OFMC verification with one session\;
	$P_3 \gets$ ProVerif tasks\;
	$P_4 \gets$ Lowest priority: OFMC with multiple sessions\;
	
	\textbf{Input:} $max\_parallel\_tasks$ \\
	Initialise $current\_running\_tasks \gets 0$\;
	
	\While{unassigned tasks list is not empty or $current\_running\_tasks > 0$}{
		\ForEach{$t$ in unassigned tasks list}{
			\eIf{$t$ is of type compiling}{
				Assign priority $P_1$ to $t$\;
				Insert $t$ into $Q$ with priority $P_1$\;
			}{
				\eIf{$t$ is of type OFMC with one session}{
					Assign priority $P_2$ to $t$\;
					Insert $t$ into $Q$ with priority $P_2$\;
				}{
					\eIf{$t$ is of type ProVerif}{
						Assign priority $P_3$ to $t$\;
						Insert $t$ into $Q$ with priority $P_3$\;
					}{
						Assign priority $P_4$ to task $t$\;
						Insert $t$ into $Q$ with priority $P_4$\;
					}
				}
			}
		}
		
			\While{queue $Q$ is not empty and $current\_running\_tasks < max\_parallel\_tasks$}{
				$t \gets$ Remove highest-priority task from $Q$\;
				Increment $current\_running\_tasks$\;

    		\textbf{new thread:} \Begin{

				Execute task $t$\;
                Wait for $t$'s termination\;
                Decrement $current\_running\_tasks$\;
    }
		 	}
	}
\end{algorithm}

\begin{figure}[!h]
\begin{centering}
\includegraphics[scale=0.7]{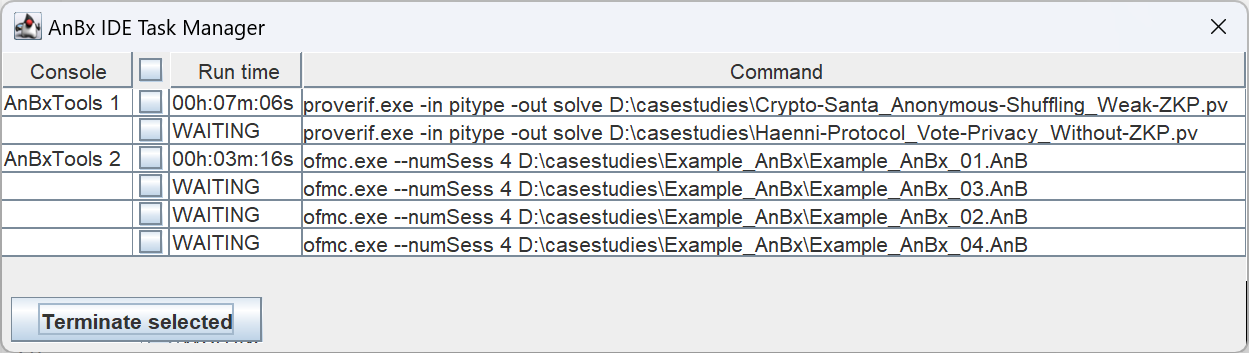}
\par\end{centering}
\caption{\label{fig:task_manager}Task manager window}
\end{figure}

To provide better control and visualisation of the running and enqueued tasks, we have also included a task manager. A button for this feature is added to the console's toolbar menu and becomes active when tasks are running. This provides a view for each task, displaying which console it has been running in, how long it has been running, whether it is waiting to start, and what its exact command line is, as presented in Figure \ref{fig:task_manager}. Users can also select any or all of the tasks and terminate them via a dedicated button.

If a task is killed, this fact is reported in the console next to its running time. This feature can be particularly useful, for example, when console outputs are logged.

The maximum number of concurrent tasks can be set in the configuration window of the \anbx{} IDE plug-in. If this number increases, new concurrent tasks can start if any are enqueued; however, no running tasks will be interrupted if it decreases. By default, it is set to the minimum of 4 and the number of available logical CPU cores.

Setting a maximum number of concurrent threads is particularly useful when verifying protocols goal by goal, typically with the \texttt{-singlegoals} option. Launching parallel processes to quickly obtain intermediate results for each goal can be very resource-intensive. Thus, one can verify several goals in parallel without consuming all resources by trying to verify all of them at once.

In the same spirit of saving resources, a global timeout in minutes can be set for the running tasks. After this time elapses, a given task will be terminated in the same manner as it would be via the task manager. The default value is 0, which disables the timeout. This feature brings convenience when launching large tasks on a cluster to avoid hogging shared resources.

\subsubsection{Attack trace reconstruction}
\label{subsec:idetracereconstruction}

We also enabled in the IDE the \anbx{} Compiler feature  \cite{GARCIA2024100038} that allows to reconstruct OFMC attack traces back into \anb{} and to generate a Java implementation of a given trace. The IDE automates the intermediate steps necessary for the reconstruction. When \textit{Reconstruct OFMC Attack Trace} is checked, the following steps ensue:

\begin{itemize}
	\item The \anbx{} protocol is exported to \anb{}.
	\item OFMC is executed on the generated \anb{}.
	\item If an attack is found, the compiler is called again on the \anbx{} protocol, using the trace for interpretation.
\end{itemize}

After this last step, the reconstructed \anb{} narration of the attack is created, and a Java implementation will also be generated if the export is set to Java. The attack narration can also be printed as \emph{ExecNarr}, which serves as an intermediate format for the concrete implementation of the attack.

With this, in addition to a UI for existing protocols, the user can see concrete results in Java emerging from an abstract attack trace, as well as a trace reconstruction expressed in \anb{}. This all happens without the user having to worry about intermediate steps, thereby providing both abstract and concrete attack artefacts within the same window with just a few clicks.
This is the second visualisation option the IDE offers to complement the Eclipse view previously mentioned in Section \ref{subsec:ideverifresultsview}.

\subsection{Environment customisation and other functionalities} \label{subsec:ide-environment}

There are several options for customisation of the environment that can improve the user experience. Configuration options are available in the \anbx{} IDE, accessible via the \texttt{AnBx Tools$\rightarrow$Configuration} menu in Eclipse. The configuration window allows users to specify paths for tool executables, configuration files, and working directories, with an automatic checking process to ensure these paths are valid and accessible. The IDE also checks file system permissions to prevent issues during operation, which can be disabled if users encounter false positives. Additionally, the window provides links for downloading tools and accessing documentation, facilitating a smooth user experience. 

Finally, the IDE tracks the versions of AnBxC, OFMC, and ProVerif, offering automatic update checks and links for manual updates to ensure users have the latest software versions. In the same dialog, the user can set the maximum number of parallel threads, the timeout value, and colouring options for the console output, already discussed in Sections \ref{subsec:consoleoutput} and \ref{subsec:taskmanager}. 

Finally, to complement and support the global user experience, we include accesses to documentation and help. It points to the respective tools' webpages and tutorials, including our own step-by-step learning tasks, with instructions\footnote{\url{https://paolo.science/anbxtutorial/}} both for IDE and command line uses, and the IDE update site \footnote{\url{https://www.dais.unive.it/~modesti/anbx/ide/}}.

\section{Evaluation}
\label{sec:ideevaluation}

In our evaluation strategy, we firstly focus on overcoming core challenges and barriers to formal methods adoption, such as complexity, accessibility, and usability, in light of how the IDE simplifies the use of verification tools. Secondly, we conduct user evaluations to assess the tool's effectiveness and user satisfaction, gathering feedback from users in educational setting. Those reviews added to usage statistics provide empirical evidence of the IDE's adoption, supporting claims of its practical impact made in Section \ref{subsec:ideaddressingchallenges}.

In Section \ref{sec:idecontribution}, we already considered performance metrics, particularly regarding single-goals versus all-goals verification times  (\ref{sec:idesinglegoals}), to demonstrate the IDE's efficiency improvements in cryptographic protocol verification tasks.  Additionally, we have shown how our IDE helps users avoid common cryptographic misconceptions, such as those related to key usage and encryption mechanisms, by offering intelligent editing features that provide immediate feedback and guidance (\ref{subsec:ideediting}).

\subsection{Addressing core challenges for formal methods adoption}
\label{subsec:ideaddressingchallenges}

We detail here how the \anbx{} IDE addresses core concerns presented in Section \ref{subsec:idechallenges} to make formal methods tools more accessible, as advocated in \cite{DBLP:conf/fmics/GaravelBP20,DBLP:journals/fac/KulikDLMSTW22,DBLP:journals/corr/LeinoW14,Davis2013}.

\begin{enumerate}
    \item \textit{Complexity:} The IDE simplifies the use of complex tools that can be employed to formalise models through high-level and standard languages. The user experience (UX) of the IDE is designed to allow manipulation of tools without requiring an in-depth understanding of the underlying theories or the precise workings of each tool. Overall, the IDE reduces the usability barrier and facilitates the work of novice users who are unfamiliar with formal methods tools and techniques, with general editing support (Section \ref{subsec:ideediting}).
    
    \item \textit{Limited Tool Integration:} We identify and integrate relevant state-of-the-art tools, providing the user with the option to run all of them from a single protocol standpoint, achieving an end-to-end workflow. This aligns with the demand for the consolidation of existing tools and methods \cite{DBLP:conf/fmics/GaravelBP20}.
    
    \item \textit{Unfamiliar Interfaces:} We integrate several tools that have different and very specific command-line interfaces. The IDE offers a uniform access point to these tools in a common Eclipse environment for consistency, limiting back-and-forth actions and making them more accessible to the public without hindering more specialised usage.

    \item \textit{Interpretability of Results:} Several features of the IDE are designed to improve the way tool results are presented to the user, making them easier to interpret: console colouring (\ref{subsec:consoleoutput}), verification result views (\ref{subsec:ideverifresultsview}), single-goal handling (\ref{sec:idesinglegoals}), and attack trace reconstruction (\ref{subsec:idetracereconstruction}). 

    \item \textit{Scalability Issues:} By supporting two verification tools, we can overcome some individual limitations and provide comparative results for a finer understanding of verification. The task manager, scheduler, concurrent threads cap, and timeout  (Section \ref{subsec:taskmanager}) help to prevent resource exhaustion and confusion with numerous running tasks.
    
    \item \textit{Limited Documentation:} We provide immediate access to documentation about the tools, reducing the need to search for it (Figure \ref{fig:anbxcdialog}). Overall, the UX of the IDE minimises the need to consult documentation in order to perform tasks.
\end{enumerate}

In summary, by incorporating user-friendly features such as editing support, automated error reporting, and intuitive workflows, the
\anbx{} IDE addresses core concerns raised in \cite{DBLP:conf/fmics/GaravelBP20} (Section \ref{subsec:idechallenges}) to make formal methods tools more accessible, enabling a broader user base in the field of the design and analysis of security protocols and their implementation.

\subsection{User Evaluation}
\label{subsec:user-evaluation}

To assess the potential benefits of our IDE in monitoring the effectiveness of our teaching activities, we surveyed a group of Cybersecurity students at Teesside University between 2023 and 2024 about their experience using the environment and associated tools. We invited 51 students that we identified as having potentially used the toolkit at some point in their studies at our university, and received responses from 35 of them (response rate: 68.8\%): 30 postgraduate and 5 undergraduate students, comprising 8 female and 27 male participants. It should be noted that almost all individuals in this sample differ from those who participated in the assessment of cryptographic misconceptions (Section \ref{subsec:survey-misconceptions}). However, the postgraduate students’ backgrounds are similar to those previously discussed, while the undergraduate participants included a mix of home and international students.

The survey was structured into sections to capture participants' experiences with the \anbx{} tools and their technical backgrounds. It employed a 5-point Likert scale \cite{likert1932technique,clark2019constructing} for most questions, allowing participants to rate the usefulness of different functionalities and their own technical proficiency. The following areas were covered:
\begin{itemize}[itemsep=0.3mm]
	\item Operating system usage
	\item Types of projects where the tools were used
	\item Usefulness of specific IDE functionalities
	\item Overall ratings of the IDE (functionality, stability, performance, usability)
	\item Importance of the tools in completing their projects
	\item Frequency of usage
	\item Technical background in computing, programming, cybersecurity, cryptography, and formal/mathematical methods for security
\end{itemize}

In addition to these structured questions, we asked about participants' interest in using specific tools in the future. We also provided open-ended questions that allowed them to give more detailed feedback on their experiences and suggestions for improvement.

We used LimeSurvey \cite{limesurvey} to collect the responses, and in all cases, the survey was administered at least one month after they had received their grades, allowing for a more objective evaluation of the IDE’s usefulness. 

Here is an overview of the results. While reading the results, it should be considered that the option \textit{Never used this function} has not been discounted from the calculation of how useful the participants found the features.\\

\noindent\textbf{Operating System Usage}
Multiple responses were allowed. The majority of participants used Windows (94.29\%), with a small proportion using Linux (14.29\%) and MacOS (8.57\%).

\noindent\textbf{Project Types}
Most participants used the IDE and supported tools for their MSc projects (60\%), while 31.43\% used it for advanced practice activities, a project-based module taken by MSc students. The remaining participants employed it for other university work (e.g., lab activities, other projects).

\noindent\textbf{IDE Functionality}
Participants rated several functionalities, with the following response options: \textit{Very useful}, \textit{Somewhat useful}, \textit{Not useful}, \textit{Not useful at all}, and \textit{Never used this function}. The results show that:
\begin{itemize}
	\item 68.57\% found the visualisation and editing of \anbx/\anb~files \textit{Very useful}, 17.14\% \textit{Somewhat useful}, 11.43\% \textit{Never used this function}.
	\item 54.29\% found the same functionality \textit{Very useful} for ProVerif files,  17.14\% \textit{Somewhat useful}, with 25.71\% reporting never using it.
	\item The generation of Java code with the \anbx~Compiler was highly appreciated, with 88.57\% finding it \textit{Very useful}, with just 2.86\% having never used it.
	\item The generation of \anb~and ProVerif files also had high utility, with 82.86\% and 60\% of participants respectively, rating it \textit{Very useful}. Again, the \textit{Never used this function} was higher for ProVerif (22.86\%) than \anb~(5.71\%).
	\item 77.14\% found \textit{Very useful} the IDE support for running the generated Java code, with just 8.57\% not having used the feature.
\end{itemize}

\noindent\textbf{Verification Tasks} Verification tasks were among the most common features utilised by the sampled population.
\begin{itemize}
	\item 62.86\% found OFMC verification of \anb~files \textit{Very useful}, and 22.86\% found it \textit{Somewhat useful}. 5.71\% \textit{Never used this function}.
	\item 40\% found ProVerif verification \textit{Very useful}, with 31.43\% finding it \textit{Somewhat useful}. 22.86\% \textit{Never used this function}.
	\item Single goal generation for ProVerif and \anb{} files has been described as useful for those who used the functionality, though it had lower usage compared to other features: 40\% \textit{Very useful} for ProVerif, 51.43\% for \anb. 22.86\% \textit{Never used this function} for \anb, 37.14\% for ProVerif.
\end{itemize}
\textbf{Logging, Task Monitoring, and Configuration} These features attracted high interests and they were considered overall very useful.
\begin{itemize}
	\item Tool output logging was considered \textit{Very useful} by 71.43\% of respondents, as it allows to archive results for further analysis. 8.57\% \textit{Never used this function}.
	\item Monitoring task progress (verification, code generation) was rated \textit{Very useful} by 74.29\%, with only 5.71\% never using this functionality.
	\item Configuring code generation parameters and concurrent task numbers was also appreciated, with 71.43\% finding configuration \textit{Very useful}. 8.57\% \textit{Never used this function}.
\end{itemize}

\noindent\textbf{General Tool Rating}
Overall, participants rated the \anbx{} IDE highly in terms of user experience. The options for the responses were: \textit{Very good}, \textit{Good}, \textit{Acceptable}, \textit{Poor}, and \textit{Very poor}.

\begin{itemize}
	\item \textit{Functionality}: 57.14\% rated it \textit{Very good}, 37.14\% \textit{Good}.
	\item \textit{Stability}: 40\% rated it \textit{Very good}, and 48.57\% \textit{Good}.
	\item \textit{Performance}: 57.14\% found the performance \textit{Very good}, and 34.29\% rated it \textit{Good}.
	\item \textit{Usability}: 45.71\% rated usability \textit{Very good}, and 40\% \textit{Good}.
\end{itemize}

\noindent\textbf{Importance of Tools in participants' activities}
We also asked how important the tools were in order to complete the technical part of their projects or activities. The options offered for the responses were:

\begin{itemize}
	\item 65.71\% of students considered the tools \textit{Very important} for completing their projects, and 20\% rated them \textit{Important}.
	\item Regarding usage frequency, 57.14\% used the tools \textit{Most of the time}, and 22.86\% used them \textit{Often} during their projects.
\end{itemize}

\noindent\textbf{Technical Background}
For the self-assessment of technical skills, we offered the following options: \textit{High}, \textit{Moderately high}, \textit{Average}, \textit{Low}, and \textit{None}. The results showed that:
\begin{itemize}
	\item 48.57\% rated themselves \textit{Moderately high} in computing, 28.57\% \textit{Average} and 22.86\% \textit{High}.
	\item 40\% rated themselves \textit{Average} in programming, with 25.71\% having low programming skills. Only 8.57\% rated themselves \textit{High}.
	\item 51.43\% rated themselves \textit{Moderately high} in cybersecurity, with 31.43\% having \textit{High} proficiency.
	\item Cryptography knowledge was more varied: 14.29\% \textit{High}, 31.43\% \textit{Moderately high}, and 42.86\% \textit{Average}.
	\item Formal/mathematical methods for security were less familiar, with only 17.14\% having \textit{High} proficiency, 28.57\% \textit{Moderately high}, 37.14\% \textit{Average} and 11.43\% reporting no experience.
\end{itemize}

The survey results demonstrate that the \anbx{} tools and IDE are regarded as highly useful, especially for Java code generation and \anbx{} file manipulation. Participants appreciated the ability to monitor tasks and ure verification processes, highlighting the importance of these features for security protocol development. Windows was the dominant operating system used, and the tools were primarily employed for MSc projects and advanced practice. 

The average general feedback on functionality, performance, and usability was largely positive (90\%), but the fact that the \textit{Very good} and \textit{Good} ratings are evenly split indicates that some improvements may be considered to enhance further the user experience.

Some results were surprising, particularly from the self-assessment survey. More than a quarter of the students described their programming skills as low at the master's level, and few claimed high proficiency. In contrast, they reported having more knowledge of cryptography and formal methods than of programming. The self-assessment for general cybersecurity was also very favourable, with more than 80\% indicating they had \textit{High} or \textit{Moderately high} mastery. Being enrolled in a security-related course likely contributed to their feeling of familiarity with security and verification concepts. 

The discrepancies between their self-assessment and the results obtained from the knowledge survey detailed in Section \ref{subsec:survey-misconceptions} highlight their lack of concrete experience with subtle security issues, which their projects are designed to start addressing. Given their misconceptions, user assistance was crucial, explaining why a large majority rated the IDE as an essential tool for completing their activities.

Our findings are a continuation of our past research on the pertinence of integrating formal methods in security education \cite{modesti_2020}. In this work, we reported that high-level abstractions enable students with limited security and programming knowledge to implement simple secure communication applications. This showed that it was possible to demystify formal concepts for individuals without PhD-level knowledge by leveraging symbolic modelling. Students acknowledged the necessity of implementing security practices at early development stages, and feedback on this IDE indicates that we provide viable support in this regard.

Notably, all the students surveyed in this evaluation that used the toolkit in individual master's projects successfully completed their assessment with an average mark of 67/100, which is substantially in line with the results achieved by their peers undertaking other master's projects. However, it is not possible to perform a quantitative evaluation, as projects are very different in nature, and we cannot directly compare our findings with similar activities conducted using different methodologies. Moreover, for pedagogical reasons, we could not enforce the use of particular tools on students, as their selection was part of the project proposal. The students played a leading role in choosing a particular technology, akin to real-world scenarios. We provided guidance but allowed them the freedom to reflect on alternative solutions.

All other students included in the sample also successfully completed their assessments, but their results were either pass/fail, or the specific mark was not significantly affected by the use of the tools.

The diverse technical backgrounds and varying levels of proficiency in cryptography and formal methods suggest that the IDE could support users with different levels of expertise, possibly through more guided tutorials or enhanced error messages. By addressing these aspects, the IDE can continue to effectively support both novice and advanced users in security protocol development.

Moreover, considering comments and discussions with the students, we understand that there is room to enhance usability, and based on this feedback, we fixed some bugs and improved the overall user experience.

Looking ahead, more than two-thirds of the participants stated that they would definitely use the toolkit if applicable to their future projects. None responded that they would probably not or definitely not use it in such settings.

\subsection{Usage Statistics}

\begin{figure}[h!]
  \centering
\begin{tikzpicture}
\begin{axis}[
    ybar , ymin=0,
    width=\textwidth*0.75,
    height=0.5\textwidth,
    ylabel=Installations,
    ytick = {0, 25, 50,75,100,125,150},
    xlabel=Year,
    xticklabel style={rotate=90,anchor=near xticklabel},
    xticklabels from table={idestats.csv}{Year},
    xtick=data,
    xticklabel style = {font=\small},
    legend cell align=left,
    legend style={at={(0.02,0.78)},anchor=south west,font=\scriptsize}]
    \addplot table [x expr=\coordindex, y=Install]{idestats.csv};
    \end{axis}
\end{tikzpicture}
 \caption{Eclipse Marketplace statistics for the \anbx{} IDE (01/2018 -- 10/2024)}
  \label{fig:idestats}
\end{figure}
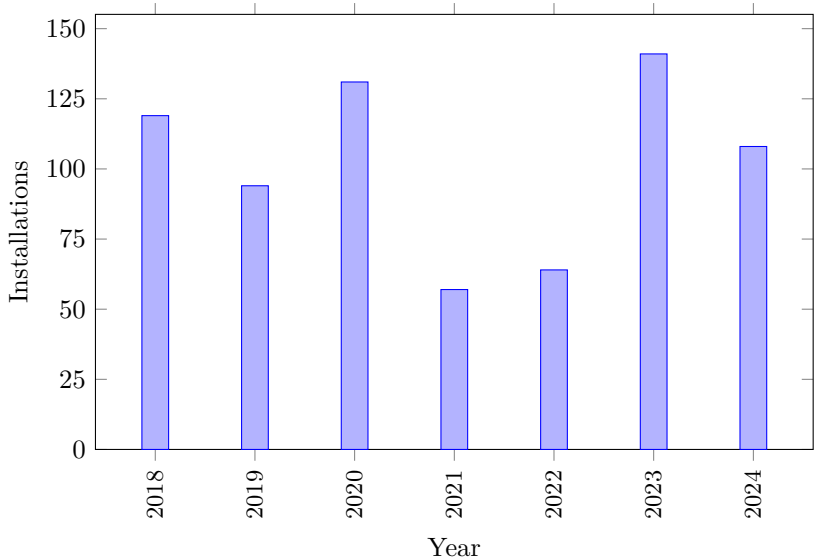

To look beyond the individuals we surveyed, we also examined the Eclipse Marketplace statistics, synthesised in Figure \ref{fig:idestats}. The first preliminary version of the tool was published in November 2017, so the data ranges from 2018 to October 2024. We observe a significant installation rate, around 100 installations per year. This demonstrates that this IDE has generated interest for years, attracting a broader community than just our group and the survey participants. It should be noted that the Eclipse Marketplace is not the only place where the installation process can begin, as it is also possible to install the plug-in directly from the distribution/update website run by the authors.

We also have information about the downloads of platform specific back-end tools used by the IDE for the period from June 2021 to October 2024, broken down by operating system: \textit{Windows} (509 downloads, 72\%), \textit{Linux} (103 downloads, 14.5\%), and \textit{macOS} (97 downloads, 13.5\%). Anecdotally, we have evidence of downloads from different countries such as Germany, Poland, the United Kingdom, the United States, China, Italy, France, Spain, Russia, India, Bulgaria, and  Denmark, among others. 

\subsection{Assumptions and Limitations}

The analysis and evaluation of the IDE for security protocol development were based on the following assumptions. First, we assumed that users, primarily postgraduate cybersecurity students, would have a basic knowledge of applied cryptography and, more generally, of security concepts, enabling them to engage with simple formal verification methods. This was tested through the initial knowledge survey (Section~\ref{subsec:survey-misconceptions}) and user evaluation (Section~\ref{subsec:user-evaluation}, technical background). We also assumed that users' cryptographic misconceptions, identified in Section~\ref{subsec:survey-misconceptions}, could be addressed through hands-on interaction with the IDE’s verification and error-handling features. This, indirectly, relies on the accuracy of self-assessment surveys in capturing users' perceptions of the IDE’s usability and educational impact (Section~\ref{subsec:user-evaluation}). 

In terms of generalising the results, we assumed that these findings could extend to professional contexts where formal methods are applied. Finally, our analysis was limited to specific tools, namely OFMC and ProVerif, as representative examples of automated symbolic verification for security protocols.

\section{Related Work}
\label{sec:iderelatedwork}

The development of fully-fledged Integrated Development Environments tailored for formal methods is not a new concept. Our contribution focuses on practicality for lowering the adoption barrier of formal methods, as promoted by Kulik \textit{et al.} \cite{DBLP:journals/fac/KulikDLMSTW22}. This is exemplified by works like Leino \textit{et al.}, who present an IDE \cite{DBLP:journals/corr/LeinoW14} for the Dafny \cite{DBLP:conf/lpar/Leino10} language, verifier, and proof assistant. The authors find, as we do, that feedback on the specification and verification results is particularly valuable.

Other examples of IDEs for formal methods tools include the Eclipse plug-in SpinRCP \cite{DBLP:conf/spin/BrezocnikVV14} for the Spin model checker \cite{DBLP:books/daglib/0020982}, which, on top of editing support, allows users to run verification and simulation tasks, with graphical visualisation and textual exportation of simulation output. A VSCode IDE \cite{DBLP:journals/corr/abs-1912-10632} exists for the Prototype Verification System \cite{DBLP:conf/cade/OwreRS92} used in avionics, with support for the evaluation of executable specifications and visualisation of proof trees.

However, when it comes to security protocols, such comprehensive environments are much rarer, and keeping them up to date is challenging. While some existing tools offer partial support, particularly at the level of editing security protocol notations like \anb{} and ProVerif, the availability of IDEs that fully integrate verification and make results more user-friendly and interpretable is limited. As per the criteria we derived from \cite{DBLP:conf/fmics/GaravelBP20,Davis2013,DBLP:journals/fac/KulikDLMSTW22} in Section \ref{subsec:idechallenges}, we summarise some contributions below.

Regarding complexity, simple \anb{}-like notations receive support from extensions or standalone editors. For example, syntax highlighting is provided for Emacs, Notepad++, Sublime Text, and TextMate via the OFMC distribution. ProVerif and Tamarin also have modes for Emacs and Vim, shipped with their respective distributions. Another project, MetaCP \cite{DBLP:journals/corr/abs-2105-09150}, allows for graphical editing of protocols in an \anb{} style on a dedicated website and offers export options to Tamarin, ProVerif, LaTeX, or C++. The Kant specification language \cite{DBLP:conf/modelsward/BraghinLRB24}, which uses an \anb{} notation, has a dedicated extension for VSCode, supporting parsing, highlighting, and semantic validation. Moreover, SPAN \cite{glouche2006security,boichut2007using} is a security protocol animator tool for AVISPA \cite{armando2005avispa} supporting HLPSL and CAS+ specifications. CAS+ is a light evolution of CASRUL \cite{DBLP:journals/entcs/Rusinowitch03}, a language in the \anb{} style.

On tool integration limitations, some editors offer the option to run the underlying tool directly from them, such as a Vim mode for ProVerif \cite{vimmodeformal}, which supports ProVerif editing as well as launching and stopping tasks. It includes similar features for other formal languages, though the ProVerif part has not been updated since 2021. A Sublime Text plug-in \cite{tamarinsublimetext} can also run tasks for Tamarin as a workflow aid, last updated in 2020.

To leverage simple, popular, and widely used interfaces, Malladi and Blanchet supplied a minimal web interface \cite{pvweb} to load protocols and launch verification tasks, and Nicolas and Cheval developed a ProVerif extension for VSCode \cite{pvvscode}. This extension is in sync with the current grammar of ProVerif as of 2024 and only provides syntax highlighting. Kant \cite{DBLP:conf/modelsward/BraghinLRB24}, being integrated into a VSCode environment, is also in a position to reach a wide audience. A different approach is taken by SPAN \cite{glouche2006security} which uses a custom GUI  and provides a virtual machine with all the required back-end tools pre-installed.

For the interpretability of results, a dedicated editor with highlighting named OFMC-GUI \cite{ofmcgui} exists, which can launch OFMC and display an attack trace as an easily readable sequence chart, but it has been unmaintained since 2020. A standalone ProVerif editor \cite{pveditor} has also been implemented in Python, providing highlighting, error reporting, an outline, and a verification summary. However, it has not seen releases since 2013. On a similar level of abstraction as ProVerif, Tamarin \cite{tamarin2012automated} can express its attack traces in the form of a graph of states violating security constraints.

As these tools usually focus on one language only, choosing different formalisms and levels of abstraction to better suit a broad range of problems is more complicated and requires multiple toolsets.

Access to documentation is also limited, often requiring users to read terminal output or visit the tool's webpage. While necessary for detailed work requiring comprehensive tutorials and examples, quick GUI-assisted documentation retrieval is usually absent from these contributions.\\

It appears that other tools in the literature often stop short of offering comprehensive support for tasks from editing to verification and interpretation of results, in a popular Eclipse environment for security protocols. The \anbx{} IDE integrates the entire process of modelling, verification, and implementation of security protocols in a way that makes the tools’ output more accessible and understandable. This includes advanced features such as console output colouring, task management, and result visualisation, which help users better interpret complex verification outcomes. This interpretation is a common challenge among users unfamiliar with the underlying formal models.

To our knowledge, the \anbx{} IDE is currently the only editor that supports the entire workflow of editing, verification, implementation, and code generation for \anbx{}, OFMC \anb{}, and ProVerif notations. We support two main back-end tools and formalisms, and the IDE parses and logs their outputs for user visualisation and analysis. Additionally, documentation is immediately available, and the latest supported versions are advertised to users to allow them to benefit from the most recent tool features. A detailed user evaluation is also rare for security protocol handling in IDEs, an aspect that we cover in this paper.

\section{Conclusion and Future Work}
\label{sec:ideconclusion}

In this paper, we introduce an Integrated Development Environment (IDE) specifically designed to support the formal development, verification, and implementation of security protocols. This research provides significant outcomes for both practical and educational applications.

First, the IDE’s intuitive interface integrates formal verification tools with features like error-checking and visual feedback, making formal methods more accessible. It reduces the traditionally steep learning curve associated with formal verification~\cite{DBLP:conf/fmics/GaravelBP20, DBLP:journals/fac/KulikDLMSTW22}, enabling even non-expert users to engage in secure and effective protocol design.

Second, by incorporating the \anb{} notation, \anbx{} Compiler, OFMC, and ProVerif tools within a single environment, the IDE streamlines the workflow. This automation reduces the complexity and time needed for protocol verification, allowing both developers and students to focus on design accuracy without dealing with low-level configuration tasks.

For students and early-career professionals, the \anbx{} IDE and its support documentation, can be also a valuable educational support tool. It simplifies the formalisation of security protocols and promotes a deeper understanding of cybersecurity principles. By clarifying common cryptographic misconceptions, the IDE helps learners to develop secure protocol design skills.

Additionally, with its built-in support at each stage of development, the IDE promotes a proactive approach to cybersecurity, potentially reducing the risks associated with the deployment of vulnerable protocols. Practitioners who aim to incorporate formal verification into the software lifecycle may streamline the process in practical contexts.

In summary, this research outcome bridges the gap between theoretical formal methods and practical, accessible tools, addressing challenges highlighted in the literature, such as complexity, limited tool integration, unfamiliar interfaces, interpretability of results, scalability issues, and sparse documentation. The IDE’s effectiveness demonstrates that these barriers to formal verification can be lowered. 

Although the IDE currently integrates tools like OFMC and ProVerif, its flexible approach makes it adaptable to other formal languages, verifiers, and IDEs. This adaptability means that core features like task automation, result interpretation, and error handling could be implemented across a broader spectrum of formal methods applications, enhancing the IDE’s relevance.

The IDE was developed using the popular Eclipse developer platform, supporting a wide range of programming languages, both mainstream and domain-specific. This approach enhances the tool’s applicability beyond our educational community, and the IDE has been distributed through the Eclipse Marketplace to maximise accessibility.

Adding new languages and tools to the IDE, Tamarin \cite{tamarin2012automated,DBLP:conf/cav/MeierSCB13} for example, would be an interesting way forward. As time and reviews go on, we plan to continue adding quality-of-life features and extend our toolset's reach.

Other future work may include evaluating how effectively various languages for security protocol specification capture and support the designer's mental models. This assessment could help identify more precisely ways to reduce errors and misconceptions.

\paragraph{Acknowledgements}
The authors express their gratitude to Leo Freitas for engaging in constructive discussions, Sebastian Mödersheim for his support with OFMC, and Bruno Blanchet and Vincent Cheval for their support with ProVerif. They also thank the users, particularly the students of Teesside University, for their feedback and suggestions regarding the \anbx{} IDE and related tools.

\bibliographystyle{elsarticle-num}
\bibliography{literature}

\begin{thebibliography}{100}
\expandafter\ifx\csname url\endcsname\relax
  \def\url#1{\texttt{#1}}\fi
\expandafter\ifx\csname urlprefix\endcsname\relax\def\urlprefix{URL }\fi
\expandafter\ifx\csname href\endcsname\relax
  \def\href#1#2{#2} \def\path#1{#1}\fi

\bibitem{anbx-ide-2024}
R.~Garcia, P.~Modesti, A practical approach to formal methods: {An Eclipse}
  integrated development environment ({IDE}) for security protocols,
  Electronics 13~(23) (2024).
\newblock \href {https://doi.org/10.3390/electronics13234660}
  {\path{doi:10.3390/electronics13234660}}.

\bibitem{DBLP:conf/ccs/VanhoefP17}
M.~Vanhoef, F.~Piessens, Key reinstallation attacks: Forcing nonce reuse in
  {WPA2}, in: B.~Thuraisingham, D.~Evans, T.~Malkin, D.~Xu (Eds.), Proceedings
  of the 2017 {ACM} {SIGSAC} Conference on Computer and Communications
  Security, {CCS} 2017, Dallas, TX, USA, October 30 - November 03, 2017, {ACM},
  2017, pp. 1313--1328.
\newblock \href {https://doi.org/10.1145/3133956.3134027}
  {\path{doi:10.1145/3133956.3134027}}.

\bibitem{heartbleed2014}
S.~Cassidy, Diagnosis of the {OpenSSL Heartbleed Bug}, available online:
  \url{https://www.seancassidy.me/diagnosis-of-the-openssl-heartbleed-bug.html}
  (accessed on 22 November 2024) (2014).

\bibitem{Fogel2016}
B.~Fogel, S.~Farmer, H.~Alkofahi, A.~Skjellum, M.~Hafiz, POODLEs, More POODLEs,
  FREAK Attacks Too: How Server Administrators Responded to Three Serious Web
  Vulnerabilities, Springer International Publishing, 2016, pp. 122--137.
\newblock \href {https://doi.org/10.1007/978-3-319-30806-7_8}
  {\path{doi:10.1007/978-3-319-30806-7_8}}.

\bibitem{Alashwali2018}
E.~S. Alashwali, K.~Rasmussen, What’s in a Downgrade? A Taxonomy of Downgrade
  Attacks in the TLS Protocol and Application Protocols Using TLS, Springer
  International Publishing, 2018, pp. 468--487.
\newblock \href {https://doi.org/10.1007/978-3-030-01704-0_27}
  {\path{doi:10.1007/978-3-030-01704-0_27}}.

\bibitem{Cremers2016}
C.~Cremers, M.~Horvat, S.~Scott, T.~van~der Merwe, Automated analysis and
  verification of tls 1.3: 0-rtt, resumption and delayed authentication, in:
  2016 IEEE Symposium on Security and Privacy (SP), IEEE, 2016.
\newblock \href {https://doi.org/10.1109/sp.2016.35}
  {\path{doi:10.1109/sp.2016.35}}.

\bibitem{Bhargavan2017}
K.~Bhargavan, B.~Blanchet, N.~Kobeissi, Verified models and reference
  implementations for the {TLS} 1.3 standard candidate, in: 2017 {IEEE}
  Symposium on Security and Privacy, {SP} 2017, San Jose, CA, USA, May 22-26,
  2017, {IEEE} Computer Society, 2017, pp. 483--502.
\newblock \href {https://doi.org/10.1109/SP.2017.26}
  {\path{doi:10.1109/SP.2017.26}}.

\bibitem{blanchet2018composition}
B.~Blanchet, Composition theorems for cryptoverif and application to {TLS} 1.3
  (2018) 16--30\href {https://doi.org/10.1109/csf.2018.00009}
  {\path{doi:10.1109/csf.2018.00009}}.

\bibitem{DBLP:journals/joc/Cohn-GordonCDGS20}
K.~Cohn{-}Gordon, C.~Cremers, B.~Dowling, L.~Garratt, D.~Stebila, A formal
  security analysis of the signal messaging protocol, J. Cryptol. 33~(4) (2020)
  1914--1983.
\newblock \href {https://doi.org/10.1007/S00145-020-09360-1}
  {\path{doi:10.1007/S00145-020-09360-1}}.

\bibitem{blanchet2001efficient}
B.~Blanchet, An efficient cryptographic protocol verifier based on {Prolog}
  rules, in: Computer Security Foundations Workshop, IEEE, IEEE Computer
  Society, 2001, pp. 0082--0082.
\newblock \href {https://doi.org/10.1109/csfw.2001.930138}
  {\path{doi:10.1109/csfw.2001.930138}}.

\bibitem{DBLP:conf/cav/MeierSCB13}
S.~Meier, B.~Schmidt, C.~Cremers, D.~A. Basin, The {TAMARIN} prover for the
  symbolic analysis of security protocols, in: N.~Sharygina, H.~Veith (Eds.),
  Computer Aided Verification - 25th International Conference, {CAV} 2013,
  Saint Petersburg, Russia, July 13-19, 2013. Proceedings, Vol. 8044 of Lecture
  Notes in Computer Science, Springer, 2013, pp. 696--701.
\newblock \href {https://doi.org/10.1007/978-3-642-39799-8_48}
  {\path{doi:10.1007/978-3-642-39799-8_48}}.

\bibitem{basin2005ofmc}
D.~Basin, S.~M\"odersheim, L.~Vigan\`o, {OFMC}: A symbolic model checker for
  security protocols, International Journal of Information Security 4~(3)
  (2005) 181--208.
\newblock \href {https://doi.org/10.1007/s10207-004-0055-7}
  {\path{doi:10.1007/s10207-004-0055-7}}.

\bibitem{Dark2015c}
M.~Dark, S.~Belcher, M.~Bishop, I.~Ngambeki, Practice, practice, practice...
  secure programmer!, in: Proceeding of the 19th Colloquium for Information
  System Security Education, 2015.

\bibitem{CVE-2020-13777}
NIST, Cve-2020-13777, available online:
  \url{https://nvd.nist.gov/vuln/detail/CVE-2020-13777/} (accessed on 22
  November 2024) (2020).

\bibitem{DBLP:conf/fmics/GaravelBP20}
H.~Garavel, M.~H. ter Beek, J.~van~de Pol, The 2020 expert survey on formal
  methods, in: M.~H. ter Beek, D.~Nickovic (Eds.), Formal Methods for
  Industrial Critical Systems - 25th International Conference, {FMICS} 2020,
  Vienna, Austria, September 2-3, 2020, Proceedings, Vol. 12327 of Lecture
  Notes in Computer Science, Springer, 2020, pp. 3--69.
\newblock \href {https://doi.org/10.1007/978-3-030-58298-2_1}
  {\path{doi:10.1007/978-3-030-58298-2_1}}.

\bibitem{DBLP:journals/fac/KulikDLMSTW22}
T.~Kulik, B.~Dongol, P.~G. Larsen, H.~D. Macedo, S.~Schneider, P.~W.~V.
  Tran{-}J{\o}rgensen, J.~Woodcock, A survey of practical formal methods for
  security, Formal Aspects Comput. 34~(1) (2022) 1--39.
\newblock \href {https://doi.org/10.1145/3522582} {\path{doi:10.1145/3522582}}.

\bibitem{dolev83ieee}
D.~Dolev, A.~Yao, On the security of public-key protocols, IEEE Transactions on
  Information Theory 2~(29) (1983).
\newblock \href {https://doi.org/10.1109/tit.1983.1056650}
  {\path{doi:10.1109/tit.1983.1056650}}.

\bibitem{sommerville2013software}
I.~Sommerville, Software Engineering, 9th edition, Addison-Wesley, 2010.

\bibitem{bugliesi2008lbs}
M.~Bugliesi, R.~Focardi, Language based secure communication, in: Computer
  Security Foundations Symposium, 2008. CSF'08. IEEE 21st, 2008, pp. 3--16.
\newblock \href {https://doi.org/10.1109/csf.2008.17}
  {\path{doi:10.1109/csf.2008.17}}.

\bibitem{avalle2014formal}
M.~Avalle, A.~Pironti, R.~Sisto, Formal verification of security protocol
  implementations: a survey, Formal Aspects of Computing 26~(1) (2014) 99--123.
\newblock \href {https://doi.org/10.1007/s00165-012-0269-9}
  {\path{doi:10.1007/s00165-012-0269-9}}.

\bibitem{Davis2013}
J.~A. Davis, M.~Clark, D.~Cofer, A.~Fifarek, J.~Hinchman, J.~Hoffman,
  B.~Hulbert, S.~P. Miller, L.~Wagner, Study on the Barriers to the Industrial
  Adoption of Formal Methods, Springer Berlin Heidelberg, 2013, pp. 63--77.
\newblock \href {https://doi.org/10.1007/978-3-642-41010-9_5}
  {\path{doi:10.1007/978-3-642-41010-9_5}}.

\bibitem{Skevoulis2006}
S.~Skevoulis, V.~Makarov, Integrating formal methods tools into undergraduate
  computer science curriculum, in: Proceedings. Frontiers in Education. 36th
  Annual Conference, IEEE, 2006, pp. 1--6.
\newblock \href {https://doi.org/10.1109/fie.2006.322570}
  {\path{doi:10.1109/fie.2006.322570}}.

\bibitem{Scheurer2000}
T.~Scheurer, Formal Methods: The Problem Is Education, Springer Berlin
  Heidelberg, 2000, pp. 198--210.
\newblock \href {https://doi.org/10.1007/3-540-40891-6_18}
  {\path{doi:10.1007/3-540-40891-6_18}}.

\bibitem{Pomorova2016}
O.~Pomorova, S.~Lysenko, Formal and intelligent methods for security and
  resilience: Education and training issues, Information \& Security: An
  International Journal 35 (2016) 133--150.
\newblock \href {https://doi.org/10.11610/isij.3507}
  {\path{doi:10.11610/isij.3507}}.

\bibitem{DBLP:journals/ijsse/AvallePPS11}
M.~Avalle, A.~Pironti, D.~Pozza, R.~Sisto, Java{SPI}: {A} framework for
  security protocol implementation, {International Journal of Secure Software
  Engineering} 2~(4) (2011) 34--48.
\newblock \href {https://doi.org/10.4018/jsse.2011100103}
  {\path{doi:10.4018/jsse.2011100103}}.

\bibitem{anbx2015}
P.~Modesti, {AnBx}: Automatic generation and verification of security protocols
  implementations, in: 8th International Symposium on Foundations \& Practice
  of Security, Vol. 9482 of LNCS, Springer, 2015, pp. 156--173.
\newblock \href {https://doi.org/10.1007/978-3-319-30303-1_10}
  {\path{doi:10.1007/978-3-319-30303-1_10}}.

\bibitem{sps2015}
O.~Almousa, S.~M{\"o}dersheim, L.~Vigan{\`{o}}, {Alice} and {Bob}: Reconciling
  formal models and implementation, in: C.~Bodei, G.-L. Ferrari, C.~Priami
  (Eds.), Programming Languages with Applications to Biology and Security:
  Essays Dedicated to Pierpaolo Degano on the Occasion of His 65th Birthday,
  Vol. 9465 of Lecture Notes in Computer Science, Springer International
  Publishing, 2015, pp. 66--85.
\newblock \href {https://doi.org/10.1007/978-3-319-25527-9_7}
  {\path{doi:10.1007/978-3-319-25527-9_7}}.

\bibitem{wing1996lightweight}
D.~J.~J. Wing, Lightweight formal methods, H. Saiedian. An invitation to formal
  methods. IEEE Computer (1996).
\newblock \href {https://doi.org/10.1007/3-540-45251-6_1}
  {\path{doi:10.1007/3-540-45251-6_1}}.

\bibitem{DBLP:conf/enase/ZamanskySRHB18}
A.~Zamansky, M.~Spichkova, G.~Rodr{\'{\i}}guez{-}Navas, P.~Herrmann, J.~O.
  Blech, Towards classification of lightweight formal methods, in: E.~Damiani,
  G.~Spanoudakis, L.~A. Maciaszek (Eds.), Proceedings of the 13th International
  Conference on Evaluation of Novel Approaches to Software Engineering, {ENASE}
  2018, Funchal, Madeira, Portugal, March 23-24, 2018, SciTePress, 2018, pp.
  305--313.
\newblock \href {https://doi.org/10.5220/0006770803050313}
  {\path{doi:10.5220/0006770803050313}}.

\bibitem{Brucker2024}
A.~D. Brucker, D.~Marmsoler, Teaching Formal Methods in Application Domains: A
  Case Study in Computer and Network Security, Springer Nature Switzerland,
  2024, pp. 124--140.
\newblock \href {https://doi.org/10.1007/978-3-031-71379-8_8}
  {\path{doi:10.1007/978-3-031-71379-8_8}}.

\bibitem{DBLP:series/synthesis/2017Brambilla}
M.~Brambilla, J.~Cabot, M.~Wimmer, Model-Driven Software Engineering in
  Practice, Second Edition, Synthesis Lectures on Software Engineering, Morgan
  {\&} Claypool Publishers, 2017.
\newblock \href {https://doi.org/10.2200/S00751ED2V01Y201701SWE004}
  {\path{doi:10.2200/S00751ED2V01Y201701SWE004}}.

\bibitem{DBLP:journals/infsof/NguyenKKT15}
P.~H. Nguyen, M.~E. Kramer, J.~Klein, Y.~L. Traon, An extensive systematic
  review on the model-driven development of secure systems, Inf. Softw.
  Technol. 68 (2015) 62--81.
\newblock \href {https://doi.org/10.1016/J.INFSOF.2015.08.006}
  {\path{doi:10.1016/J.INFSOF.2015.08.006}}.

\bibitem{anbx-ide2017}
R.~Garcia, P.~Modesti, An {IDE} for the design, verification and implementation
  of security protocols, in: 2017 {IEEE} International Symposium on Software
  Reliability Engineering Workshops, {ISSRE} Workshops 2017, Toulouse, France,
  October 23-26, 2017, {IEEE} Computer Society, 2017, pp. 157--163.
\newblock \href {https://doi.org/10.1109/ISSREW.2017.69}
  {\path{doi:10.1109/ISSREW.2017.69}}.

\bibitem{Bettini2016}
L.~Bettini, Implementing domain-specific languages with Xtext and Xtend, Packt
  Publishing Ltd, 2016.

\bibitem{xtext2024}
{Eclipse Community}, Xtext documentation, available online:
  \url{http://eclipse.org/Xtext/documentation/} (accessed on 22 November 2024).

\bibitem{DBLP:conf/IEEEares/Modersheim09}
S.~M{\"{o}}dersheim, Algebraic properties in alice and bob notation, in:
  Proceedings of the The Forth International Conference on Availability,
  Reliability and Security, {ARES} 2009, March 16-19, 2009, Fukuoka, Japan,
  {IEEE} Computer Society, 2009, pp. 433--440.
\newblock \href {https://doi.org/10.1109/ARES.2009.95}
  {\path{doi:10.1109/ARES.2009.95}}.

\bibitem{jisa_2016}
M.~Bugliesi, S.~Calzavara, S.~Mödersheim, P.~Modesti, Security protocol
  specification and verification with anbx, Journal of Information Security and
  Applications 30 (2016) 46--63.
\newblock \href {https://doi.org/10.1016/j.jisa.2016.05.004}
  {\path{doi:10.1016/j.jisa.2016.05.004}}.

\bibitem{blanchet:hal-03366962}
B.~Blanchet, V.~Cheval, V.~Cortier, {ProVerif with Lemmas, Induction, Fast
  Subsumption, and Much More}, in: {43RD IEEE Symposium on Security and Privacy
  (S\&P'22)}, San Francisco, United States, 2022.
\newblock \href {https://doi.org/10.1109/sp46214.2022.9833653}
  {\path{doi:10.1109/sp46214.2022.9833653}}.

\bibitem{8312264}
R.~Shaukat, A.~Shahoor, A.~Urooj, {Probing into code analysis tools: A
  comparison of C\# supporting static code analyzers}, in: 2018 15th
  International Bhurban Conference on Applied Sciences and Technology (IBCAST),
  2018, pp. 455--464.
\newblock \href {https://doi.org/10.1109/IBCAST.2018.8312264}
  {\path{doi:10.1109/IBCAST.2018.8312264}}.

\bibitem{4283790}
M.~Fetaji, S.~Loskovska, B.~Fetaji, M.~Ebibi, Combining virtual learning
  environment and integrated development environment to enhance e-learning, in:
  2007 29th International Conference on Information Technology Interfaces,
  2007, pp. 319--324.
\newblock \href {https://doi.org/10.1109/ITI.2007.4283790}
  {\path{doi:10.1109/ITI.2007.4283790}}.

\bibitem{10.1145/3670795}
M.~Broy, A.~Brucker, A.~Fantechi, M.~Gleirscher, K.~Havelund, M.~A. Kuppe,
  A.~Mendes, A.~Platzer, J.~Ringert, A.~Sullivan, Does every computer scientist
  need to know formal methods?, Form. Asp. Comput.Just Accepted (Jun. 2024).
\newblock \href {https://doi.org/10.1145/3670795} {\path{doi:10.1145/3670795}}.

\bibitem{DBLP:journals/ese/GleirscherM20}
M.~Gleirscher, D.~Marmsoler, Formal methods in dependable systems engineering:
  a survey of professionals from europe and north america, Empir. Softw. Eng.
  25~(6) (2020) 4473--4546.
\newblock \href {https://doi.org/10.1007/S10664-020-09836-5}
  {\path{doi:10.1007/S10664-020-09836-5}}.

\bibitem{DBLP:books/daglib/0068766}
J.~M. Spivey, Z Notation - a reference manual {(2.} ed.), Prentice Hall
  International Series in Computer Science, Prentice Hall, 1992.

\bibitem{wadsworth1996piaget}
B.~J. Wadsworth, Piaget's Theory of Cognitive and Affective Development:
  Foundations of Constructivism, Longman Publishing, 1996.

\bibitem{bruner2009process}
J.~S. Bruner, The Process of Education, Harvard University Press, 2009.
\newblock \href {https://doi.org/10.2307/j.ctvk12qst}
  {\path{doi:10.2307/j.ctvk12qst}}.

\bibitem{DBLP:conf/etfa/RaimondoMBP23}
M.~Raimondo, S.~Marrone, S.~Bernardi, A.~Palladino, Demonstrating the necessity
  of model generation in security protocol verification, in: 28th {IEEE}
  International Conference on Emerging Technologies and Factory Automation,
  {ETFA} 2023, Sinaia, Romania, September 12-15, 2023, {IEEE}, 2023, pp. 1--8.
\newblock \href {https://doi.org/10.1109/ETFA54631.2023.10275424}
  {\path{doi:10.1109/ETFA54631.2023.10275424}}.

\bibitem{DBLP:journals/fac/PaulsonNW19}
L.~C. Paulson, T.~Nipkow, M.~Wenzel, From {LCF} to isabelle/hol, Formal Aspects
  Comput. 31~(6) (2019) 675--698.
\newblock \href {https://doi.org/10.1007/S00165-019-00492-1}
  {\path{doi:10.1007/S00165-019-00492-1}}.

\bibitem{abadi1994pep}
M.~Abadi, R.~Needham, Prudent engineering practice for cryptographic protocols,
  in: 1994 IEEE Computer Society Symposium on Research in Security and Privacy,
  1994. Proceedings., 1994, pp. 122--136.
\newblock \href {https://doi.org/10.1109/32.481513}
  {\path{doi:10.1109/32.481513}}.

\bibitem{DBLP:journals/corr/LeinoW14}
K.~R.~M. Leino, V.~W{\"{u}}stholz, The dafny integrated development
  environment, in: C.~Dubois, D.~Giannakopoulou, D.~M{\'{e}}ry (Eds.),
  Proceedings 1st Workshop on Formal Integrated Development Environment,
  {F-IDE} 2014, Grenoble, France, April 6, 2014, Vol. 149 of {EPTCS}, 2014, pp.
  3--15.
\newblock \href {https://doi.org/10.4204/EPTCS.149.2}
  {\path{doi:10.4204/EPTCS.149.2}}.

\bibitem{Unwin1999}
A.~Unwin, H.~Hofmann, Gui and command-line - conflict or synergy?, in: K.~Berk,
  M.~Pourahmadi (Eds.), Proceedings of the 31st Symposium on the Interface:
  models, predictions, and computing, Schaumburg, Illinois, June 9 - 12, 1999,
  Computing science and statistics, Interface Foundation of North America,
  1999, pp. 246--253.

\bibitem{DBLP:conf/soups/TabassumWL17}
M.~Tabassum, S.~Watson, H.~R. Lipford,
  \href{https://www.usenix.org/conference/soups2017/workshop-program/wsiw2017/tabassum}{Comparing
  educational approaches to secure programming: Tool vs. {TA}}, in: Thirteenth
  Symposium on Usable Privacy and Security, {SOUPS} 2017, Santa Clara, CA, USA,
  July 12-14, 2017, {USENIX} Association, 2017.
\newline\urlprefix\url{https://www.usenix.org/conference/soups2017/workshop-program/wsiw2017/tabassum}

\bibitem{DBLP:conf/profes/Kuusinen15}
K.~Kuusinen, Software developers as users: Developer experience of a
  cross-platform integrated development environment, in: P.~Abrahamsson,
  L.~Corral, M.~Oivo, B.~Russo (Eds.), Product-Focused Software Process
  Improvement - 16th International Conference, {PROFES} 2015, Bolzano, Italy,
  December 2-4, 2015, Proceedings, Vol. 9459 of Lecture Notes in Computer
  Science, Springer, 2015, pp. 546--552.
\newblock \href {https://doi.org/10.1007/978-3-319-26844-6_40}
  {\path{doi:10.1007/978-3-319-26844-6_40}}.

\bibitem{DBLP:conf/wipsce/LindmeierM20}
A.~Lindmeier, A.~M{\"{u}}hling, Keeping secrets: {K-12} students' understanding
  of cryptography, in: T.~Brinda, M.~Armoni (Eds.), WiPSCE '20: Workshop in
  Primary and Secondary Computing Education, Virtual Event, Germany, October
  28-30, 2020, {ACM}, 2020, pp. 14:1--14:10.
\newblock \href {https://doi.org/10.1145/3421590.3421630}
  {\path{doi:10.1145/3421590.3421630}}.

\bibitem{DBLP:conf/nordichi/Geels24}
J.~Geels, Ordinary users do not understand digital signatures, in: Proceedings
  of the 13th Nordic Conference on Human-Computer Interaction, NordiCHI 2024,
  Uppsala, Sweden, October 13-16, 2024, {ACM}, 2024, pp. 66:1--66:15.
\newblock \href {https://doi.org/10.1145/3679318.3685402}
  {\path{doi:10.1145/3679318.3685402}}.

\bibitem{DBLP:journals/tr/BragaDALV19}
A.~M. Braga, R.~Dahab, N.~Antunes, N.~Laranjeiro, M.~Vieira, Understanding how
  to use static analysis tools for detecting cryptography misuse in software,
  {IEEE} Trans. Reliab. 68~(4) (2019) 1384--1403.
\newblock \href {https://doi.org/10.1109/TR.2019.2937214}
  {\path{doi:10.1109/TR.2019.2937214}}.

\bibitem{DBLP:journals/csur/GleirscherFW20}
M.~Gleirscher, S.~Foster, J.~Woodcock, New opportunities for integrated formal
  methods, {ACM} Comput. Surv. 52~(6) (2020) 117:1--117:36.
\newblock \href {https://doi.org/10.1145/3357231} {\path{doi:10.1145/3357231}}.

\bibitem{DBLP:conf/ifm/GlabbeekHW18}
R.~J. van Glabbeek, P.~H{\"{o}}fner, D.~van~der Wal, Analysing
  awn-specifications using mcrl2 (extended abstract), in: C.~A. Furia,
  K.~Winter (Eds.), Integrated Formal Methods - 14th International Conference,
  {IFM} 2018, Maynooth, Ireland, September 5-7, 2018, Proceedings, Vol. 11023
  of Lecture Notes in Computer Science, Springer, 2018, pp. 398--418.
\newblock \href {https://doi.org/10.1007/978-3-319-98938-9_23}
  {\path{doi:10.1007/978-3-319-98938-9_23}}.

\bibitem{DBLP:conf/se/RungeSCTKW21}
T.~Runge, I.~Schaefer, L.~Cleophas, T.~Th{\"{u}}m, D.~G. Kourie, B.~W. Watson,
  Tool support for correctness-by-construction, in: A.~Koziolek, I.~Schaefer,
  C.~Seidl (Eds.), Software Engineering 2021, Fachtagung des GI-Fachbereichs
  Softwaretechnik, 22.-26. Februar 2021, Braunschweig/Virtuell, Vol. {P-310} of
  {LNI}, Gesellschaft f{\"{u}}r Informatik e.V., 2021, pp. 93--94.
\newblock \href {https://doi.org/10.18420/SE2021_34}
  {\path{doi:10.18420/SE2021_34}}.

\bibitem{DBLP:conf/facs2/FaresBF24}
E.~Fares, J.~Bodeveix, M.~Filali, Correct pattern-based development through
  refinements and weakest preconditions calculus, in: D.~Marmsoler, M.~Sun
  (Eds.), Formal Aspects of Component Software - 20th International Conference,
  {FACS} 2024, Milan, Italy, September 9-10, 2024, Proceedings, Vol. 15189 of
  Lecture Notes in Computer Science, Springer, 2024, pp. 59--78.
\newblock \href {https://doi.org/10.1007/978-3-031-71261-6_4}
  {\path{doi:10.1007/978-3-031-71261-6_4}}.

\bibitem{DBLP:conf/nfm/BonfantiCGM17}
S.~Bonfanti, M.~Carissoni, A.~Gargantini, A.~Mashkoor, Asm2c++: {A} tool for
  code generation from abstract state machines to arduino, in: C.~W. Barrett,
  M.~D. Davies, T.~Kahsai (Eds.), {NASA} Formal Methods - 9th International
  Symposium, {NFM} 2017, Moffett Field, CA, USA, May 16-18, 2017, Proceedings,
  Vol. 10227 of Lecture Notes in Computer Science, 2017, pp. 295--301.
\newblock \href {https://doi.org/10.1007/978-3-319-57288-8_21}
  {\path{doi:10.1007/978-3-319-57288-8_21}}.

\bibitem{Lowe97}
G.~Lowe, A hierarchy of authentication specifications, in: CSFW'97, IEEE
  Computer Society Press, 1997, pp. 31--43.

\bibitem{DBLP:journals/corr/GaleottiFMFZ14}
J.~P. Galeotti, C.~A. Furia, E.~May, G.~Fraser, A.~Zeller, Automating full
  functional verification of programs with loops, CoRR abs/1407.5286 (2014).
\newblock \href {http://arxiv.org/abs/1407.5286} {\path{arXiv:1407.5286}},
  \href {https://doi.org/10.48550/arxiv.1407.5286}
  {\path{doi:10.48550/arxiv.1407.5286}}.

\bibitem{DBLP:conf/sp/BarbosaBBBCLP21}
M.~Barbosa, G.~Barthe, K.~Bhargavan, B.~Blanchet, C.~Cremers, K.~Liao,
  B.~Parno, Sok: Computer-aided cryptography, in: 42nd {IEEE} Symposium on
  Security and Privacy, {SP} 2021, San Francisco, CA, USA, 24-27 May 2021,
  {IEEE}, 2021, pp. 777--795.
\newblock \href {https://doi.org/10.1109/SP40001.2021.00008}
  {\path{doi:10.1109/SP40001.2021.00008}}.

\bibitem{avispa-manual}
A.~Team, Avispa v1.0 user manual, available online:
  \url{https://people.rennes.inria.fr/Thomas.Genet/Crypt/AVISPA_manual.pdf}
  (accessed on 22 November 2024).

\bibitem{DBLP:journals/jlp/DelauneH17}
S.~Delaune, L.~Hirschi, A survey of symbolic methods for establishing
  equivalence-based properties in cryptographic protocols, Journal of Logical
  and Algebraic Methods in Programming 87 (2017) 127--144.
\newblock \href {https://doi.org/10.1016/j.jlamp.2016.10.005}
  {\path{doi:10.1016/j.jlamp.2016.10.005}}.

\bibitem{blanchetproverifmanual}
B.~Blanchet, B.~Smyth, V.~Cheval, {ProVerif} 2.05: Automatic cryptographic
  protocol verifier, user manual and tutorial, available online:
  \url{https://bblanche.gitlabpages.inria.fr/proverif/manual.pdf} (accessed on
  22 November 2024) (2023).

\bibitem{DBLP:journals/sigops/Carlsen94}
U.~Carlsen, Optimal privacy and authentication on a portable communications
  system, {ACM} {SIGOPS} Oper. Syst. Rev. 28~(3) (1994) 16--23.
\newblock \href {https://doi.org/10.1145/182110.182112}
  {\path{doi:10.1145/182110.182112}}.

\bibitem{ofmcsite}
{Sebastian M{\"{o}}dersheim}, {OFMC distribution and tutorials}, available
  online: \url{https://www2.compute.dtu.dk/~samo/} (accessed on 22 November
  2024).

\bibitem{h530}
{ITU-T Recommendation H.530: Symmetric Security Procedures for H.510 (Mobility
  for H.323 Multimedia Systems and Services)} (2002).

\bibitem{kaufman2005internet}
C.~Kaufman, Internet key exchange ({IKEv2}) protocol, Tech. rep. (2005).

\bibitem{isosymkey}
S.~International Organization~for Standardization, Gen\`eve, {ISO/IEC
  9798-2:2008, Information technology – Security techniques – Entity
  Authentication – Part 2: Mechanisms using symmetric encipherment
  algorithms, Third edition} (2008).

\bibitem{isoccf}
S.~International Organization~for Standardization, Gen\`eve, {ISO/IEC
  9798-4:1999, Information technology – Security techniques – Entity
  Authentication – Part 3: Mechanisms using a cryptographic check function,
  Second edition} (1999).

\bibitem{DBLP:journals/stp/Lowe96}
G.~Lowe, Breaking and fixing the needham-schroeder public-key protocol using
  {FDR}, Softw. Concepts Tools 17~(3) (1996) 93--102.
\newblock \href {https://doi.org/10.1007/3-540-61042-1_43}
  {\path{doi:10.1007/3-540-61042-1_43}}.

\bibitem{DBLP:journals/cacm/NeedhamS78}
R.~M. Needham, M.~D. Schroeder, Using encryption for authentication in large
  networks of computers, Commun. {ACM} 21~(12) (1978) 993--999.
\newblock \href {https://doi.org/10.1145/359657.359659}
  {\path{doi:10.1145/359657.359659}}.

\bibitem{DBLP:journals/sigops/OtwayR87}
D.~J. Otway, O.~Rees, Efficient and timely mutual authentication, {ACM}
  {SIGOPS} Oper. Syst. Rev. 21~(1) (1987) 8--10.
\newblock \href {https://doi.org/10.1145/24592.24594}
  {\path{doi:10.1145/24592.24594}}.

\bibitem{DBLP:journals/tissec/Paulson99}
L.~C. Paulson, Inductive analysis of the internet protocol {TLS}, {ACM} Trans.
  Inf. Syst. Secur. 2~(3) (1999) 332--351.
\newblock \href {https://doi.org/10.1145/322510.322530}
  {\path{doi:10.1145/322510.322530}}.

\bibitem{woo1992authentication}
T.~Y. Woo, S.~S. Lam, Authentication for distributed systems, Computer 25~(1)
  (1992) 39--52.
\newblock \href {https://doi.org/10.1109/2.108052}
  {\path{doi:10.1109/2.108052}}.

\bibitem{paulson2000yahalom}
L.~C. Paulson, The yahalom protocol, in: Security Protocols: 7th International
  Workshop, Cambridge, UK, April 19-21, 1999. Proceedings 7, Springer, 2000,
  pp. 78--84.
\newblock \href {https://doi.org/10.1007/10720107_11}
  {\path{doi:10.1007/10720107_11}}.

\bibitem{GARCIA2024100038}
R.~Garcia, P.~Modesti, Automatic generation of security protocols attacks
  specifications and implementations, Cyber Security and Applications 2 (2024)
  100038.
\newblock \href {https://doi.org/10.1016/j.csa.2024.100038}
  {\path{doi:10.1016/j.csa.2024.100038}}.

\bibitem{likert1932technique}
R.~Likert, A technique for the measurement of attitudes., Archives of
  psychology (1932).

\bibitem{clark2019constructing}
L.~A. Clark, D.~Watson, Constructing validity: New developments in creating
  objective measuring instruments., Psychological assessment 31~(12) (2019)
  1412.
\newblock \href {https://doi.org/10.1037/pas0000626}
  {\path{doi:10.1037/pas0000626}}.

\bibitem{limesurvey}
{LimeSurvey Team}, Limesurvey: an open source survey tool,
  \url{https://www.limesurvey.org}, [Online; accessed 14 October 2024] (2024).

\bibitem{modesti_2020}
P.~Modesti, Integrating formal methods for security in software security
  education, Informatics in Education 19~(3) (2020) 425--454.
\newblock \href {https://doi.org/10.15388/infedu.2020.19}
  {\path{doi:10.15388/infedu.2020.19}}.

\bibitem{DBLP:conf/lpar/Leino10}
K.~R.~M. Leino, Dafny: An automatic program verifier for functional
  correctness, in: E.~M. Clarke, A.~Voronkov (Eds.), Logic for Programming,
  Artificial Intelligence, and Reasoning - 16th International Conference,
  LPAR-16, Dakar, Senegal, April 25-May 1, 2010, Revised Selected Papers, Vol.
  6355 of Lecture Notes in Computer Science, Springer, 2010, pp. 348--370.
\newblock \href {https://doi.org/10.1007/978-3-642-17511-4_20}
  {\path{doi:10.1007/978-3-642-17511-4_20}}.

\bibitem{DBLP:conf/spin/BrezocnikVV14}
Z.~Brezocnik, B.~Vlaovic, A.~Vreze, Spinrcp: the eclipse rich client platform
  integrated development environment for the spin model checker, in: N.~Rungta,
  O.~Tkachuk (Eds.), 2014 International Symposium on Model Checking of
  Software, {SPIN} 2014, Proceedings, San Jose, CA, USA, July 21-23, 2014,
  {ACM}, 2014, pp. 125--128.
\newblock \href {https://doi.org/10.1145/2632362.2632380}
  {\path{doi:10.1145/2632362.2632380}}.

\bibitem{DBLP:books/daglib/0020982}
G.~J. Holzmann, The {SPIN} Model Checker - primer and reference manual,
  Addison-Wesley, 2004.

\bibitem{DBLP:journals/corr/abs-1912-10632}
P.~Masci, C.~A. Mu{\~{n}}oz, An integrated development environment for the
  prototype verification system, in: R.~Monahan, V.~Prevosto, J.~Proen{\c{c}}a
  (Eds.), Proceedings Fifth Workshop on Formal Integrated Development
  Environment, F-IDE@FM 2019, Porto, Portugal, 7th October 2019, Vol. 310 of
  {EPTCS}, 2019, pp. 35--49.
\newblock \href {https://doi.org/10.4204/EPTCS.310.5}
  {\path{doi:10.4204/EPTCS.310.5}}.

\bibitem{DBLP:conf/cade/OwreRS92}
S.~Owre, J.~M. Rushby, N.~Shankar, {PVS:} {A} prototype verification system,
  in: D.~Kapur (Ed.), Automated Deduction - CADE-11, 11th International
  Conference on Automated Deduction, Saratoga Springs, NY, USA, June 15-18,
  1992, Proceedings, Vol. 607 of Lecture Notes in Computer Science, Springer,
  1992, pp. 748--752.
\newblock \href {https://doi.org/10.1007/3-540-55602-8_217}
  {\path{doi:10.1007/3-540-55602-8_217}}.

\bibitem{DBLP:journals/corr/abs-2105-09150}
R.~Metere, L.~Arnaboldi, Metacp: Cryptographic protocol design tool for formal
  verification, CoRR abs/2105.09150 (2021).
\newblock \href {http://arxiv.org/abs/2105.09150} {\path{arXiv:2105.09150}},
  \href {https://doi.org/10.48550/arxiv.2105.09150}
  {\path{doi:10.48550/arxiv.2105.09150}}.

\bibitem{DBLP:conf/modelsward/BraghinLRB24}
C.~Braghin, M.~Lilli, E.~Riccobene, M.~Baba, Kant: {A} domain-specific language
  for modeling security protocols, in: F.~J.~D. Mayo, L.~F. Pires, E.~Seidewitz
  (Eds.), Proceedings of the 12th International Conference on Model-Based
  Software and Systems Engineering, {MODELSWARD} 2024, Rome, Italy, February
  21-23, 2024, {SCITEPRESS}, 2024, pp. 62--73.
\newblock \href {https://doi.org/10.5220/0012386400003645}
  {\path{doi:10.5220/0012386400003645}}.

\bibitem{glouche2006security}
Y.~Glouche, T.~Genet, O.~Heen, O.~Courtay,
  \href{https://people.irisa.fr/Thomas.Genet/span/}{A security protocol
  animator tool for avispa}, in: ARTIST2 workshop on security specification and
  verification of embedded systems, Pisa, 2006, pp. 1--7.
\newline\urlprefix\url{https://people.irisa.fr/Thomas.Genet/span/}

\bibitem{boichut2007using}
Y.~Boichut, T.~Genet, Y.~Glouche, O.~Heen, Using animation to improve formal
  specifications of security protocols, in: 2nd Conference on Security in
  Network Architectures and Information Systems (SARSSI 2007), 2007, pp.
  169--182.

\bibitem{armando2005avispa}
A.~Armando, D.~Basin, Y.~Boichut, Y.~Chevalier, L.~Compagna, J.~Cu{\'e}llar,
  P.~H. Drielsma, P.-C. H{\'e}am, O.~Kouchnarenko, J.~Mantovani, et~al., The
  {AVISPA} tool for the automated validation of internet security protocols and
  applications, in: Computer Aided Verification, Springer, 2005, pp. 281--285.
\newblock \href {https://doi.org/10.1007/11513988_27}
  {\path{doi:10.1007/11513988_27}}.

\bibitem{DBLP:journals/entcs/Rusinowitch03}
M.~Rusinowitch, Automated analysis of security protocols, in: L.~Brim,
  O.~Grumberg (Eds.), 12th International Workshop on Functional and Constraint
  Logic Programming, {WFLP} 2003, in connection with RDP'03, Federated
  Conference on Rewriting, Deduction and Programming, Boulder, Colorado, USA,
  July 14, 2003, Vol.~86 of Electronic Notes in Theoretical Computer Science,
  Elsevier, 2003, pp. 12--15.
\newblock \href {https://doi.org/10.1016/S1571-0661(04)80690-X}
  {\path{doi:10.1016/S1571-0661(04)80690-X}}.

\bibitem{vimmodeformal}
Lifepillar, {Vim mode for formal languages}, available online:
  \url{https://github.com/lifepillar/vim-formal-package} (accessed on 22
  November 2024).

\bibitem{tamarinsublimetext}
B.~F. Jorden~Whitefield, Ralf~Sasse, {Sublime Text 3 plug-in for Tamarin},
  available online: \url{https://github.com/tamarin-prover/editor-sublime}
  (accessed on 22 November 2024).

\bibitem{pvweb}
S.~Malladi, B.~Blanchet, {ProVerif Web interface}, available online:
  \url{http://proverif20.paris.inria.fr/index.php} (accessed on 22 November
  2024).

\bibitem{pvvscode}
G.~Nicolas, V.~Cheval, {ProVerif Syntax Highlighting for VS Code}, available
  online:
  \url{https://marketplace.visualstudio.com/items?itemName=georgio.proverif-vscode}
  (accessed on 22 November 2024).

\bibitem{ofmcgui}
\'{U}lfur J\'{o}hann~Edvardsson, V.~J.~L. Hoffmann, {OFMC-GUI}, available
  online: \url{https://github.com/ulfur88/OFMC-GUI} (accessed on 22 November
  2024).

\bibitem{pveditor}
J.~de~Ruiter, {ProVerif Editor}, available online:
  \url{https://proverifeditor.sourceforge.net/} (accessed on 22 November 2024).

\bibitem{tamarin2012automated}
B.~Schmidt, S.~Meier, C.~Cremers, D.~Basin, Automated analysis of
  {Diffie-Hellman} protocols and advanced security properties, in: Computer
  Security Foundations Symposium (CSF), 2012 IEEE 25th, IEEE, 2012, pp. 78--94.
\newblock \href {https://doi.org/10.1109/csf.2012.25}
  {\path{doi:10.1109/csf.2012.25}}.

\end{thebibliography}

\newpage
\appendix
\section{\anbx{} IDE key features and requirements}\label{app:key-features}

The \anbx{} Integrated Development Environment (IDE) facilitates the design, verification, and implementation of security protocols, aiming to reduce the adoption barrier for formal methods tools in the security domain. Following the principles of Model-Driven Development (MDD), the environment assists users in specifying models with the simple and intuitive Alice \& Bob (\anb{}) narration language \cite{DBLP:conf/IEEEares/Modersheim09}, as well as its extension, \anbx{} \cite{anbx2015}.

Furthermore, it offers a push-button solution for the formal verification of both abstract and concrete models, along with the automatic generation of Java implementations and ProVerif specifications. The tool also supports the \textit{applied pi calculus}, which is used for modelling security protocols in ProVerif.

\subsection*{Key Features}

This \anbx{} IDE integrates with existing languages and tools for the modelling and verification of security protocols, including:
\begin{itemize}
	\item \textit{AnBx Compiler and Code Generator}: a compiler for AnBx files, and code generator \cite{anbx2015}. Targets include \textit{AnB}, \textit{ProVerif}, \textit{Java}, and \textit{Java (Docker)}.
	\item \textit{OFMC}: A model checker for security protocol verification \cite{basin2005ofmc}.
	\item \textit{ProVerif}: A verifier for cryptographic protocols \cite{blanchet2001efficient}.
\end{itemize}

\subsection*{Requirements}

To utilise the IDE, the following requirements must be met:
\begin{itemize}
	\item Java 11 or later. At the time of writing, we recommend using Java 21.
	
	\item XText Redistributable 2.34.0 or later.
	\item \textit{External Tools:} The \anbx{} Compiler, OFMC, and ProVerif must be downloaded separately. Packages for Windows, Linux, and macOS are available on the support site.
\end{itemize}

\section{Getting Started with the \anbx{} IDE}\label{app:getting-started}

This section provides a brief guide to getting started with \anbx{} IDE.

\subsection{Eclipse Plug-in Installation}

The simplest way to install the plug-in is:
\begin{enumerate}
    \item Open Eclipse and select \textit{Help} $\rightarrow$ \textit{Eclipse Marketplace...}.
    \item Search for \emph{AnBx}, then select \emph{AnBx IDE} from the list and press \textit{Install}.
    \item Ensure the update site is active for future updates by checking \textit{Help} $\rightarrow$ \textit{Install New Software...} $\rightarrow$ \textit{Manage...}, and verifying that the update site is enabled.
\end{enumerate}

Alternatively, use the Eclipse Update Manager:
\begin{itemize}
    \item Open \textit{Help} $\rightarrow$ \textit{Install New Software...}, add the following update site URL:  
    \url{https://www.dais.unive.it/~modesti/anbx/}.
    \item Follow the installation instructions provided in the Eclipse Online Help documentation.
\end{itemize}

\subsection{Back-End Tools Installation}

For a quick setup:
\begin{itemize}
    \item Download the pre-packaged back-end tools for Windows, Linux, or Mac from the AnBx website.
    \item Extract the archive to a local folder.
    \item On Mac or Linux, set execution permissions for binaries using:  
    \begin{verbatim}
    chmod u+x <filename>
    \end{verbatim}
    Alternatively, use the operating system’s GUI for setting permissions.
\end{itemize}

\subsection{Configuring the Plug-in}

To configure the plug-in:
\begin{enumerate}
    \item In Eclipse, select \textit{AnBx Tools} $\rightarrow$ \textit{Configuration}.
    \item Set the paths to:
    \begin{itemize}
        \item \textit{AnBxC Exe Path}: Path to \texttt{anbxc.exe}.
        \item \textit{Config File}: Path to \texttt{anbxc.cfg}.
        \item \textit{ProVerif Exe Path}: Path to \texttt{proverif.exe}.
        \item \textit{OFMC Exe Path}: Path to \texttt{ofmc.exe}.
    \end{itemize}
\end{enumerate}

Optionally, configure a folder for logging tool outputs via the configuration interface. Additional details are available in \ref{app:advanced-config}.

\subsection{Testing the Environment}

To verify the installation:
\begin{itemize}
    \item Open example protocols included in the \texttt{casestudies} folder from the distribution.
    \item In Eclipse, select \textit{File} $\rightarrow$ \textit{Open Project from File System...}.
\end{itemize}

Test the setup by:
\begin{enumerate}
    \item Selecting a protocol, such as \texttt{Fresh\_From\_A.AnBx}, from the \texttt{casestudies} folder.
    \item Clicking the \textit{AnBxC} button on the toolbar.
    \item Choosing an output format (e.g., \textit{AnB}, \textit{ProVerif Typed}, or \textit{Java}).
    \item Observing the verification results in the Console view.
\end{enumerate}

\subsection{Creating a New Project}

To create a new AnBx project:
\begin{enumerate}
    \item Select \textit{File} $\rightarrow$ \textit{New} $\rightarrow$ \textit{Project} $\rightarrow$ \textit{AnBx Project}.
    \item Create new AnBx/AnB files inside the \texttt{src} folder:  
    \textit{File} $\rightarrow$ \textit{New} $\rightarrow$ \textit{Other} $\rightarrow$ \textit{AnBx file}.
    \item Similarly, create ProVerif files under the \texttt{src} folder:  
    \textit{File} $\rightarrow$ \textit{New} $\rightarrow$ \textit{Other} $\rightarrow$ \textit{PV file}.
\end{enumerate}

\subsection{Running Protocol Verification and Code Generation}

The output of verification or code generation will be displayed in the \textit{Console} view. If the Console is not visible:
\begin{itemize}
    \item Select \textit{Window} $\rightarrow$ \textit{Show View} $\rightarrow$ \textit{Console}.
\end{itemize}

\section{Advanced Configuration}\label{app:advanced-config}

This section presents the configuration options of the \anbx{} IDE. The settings can be accessed through the \textit{AnBx Tools $\rightarrow$ Configuration} menu in the Eclipse toolbar.

\begin{figure}[!h]
\centering
\includegraphics[scale=0.5]{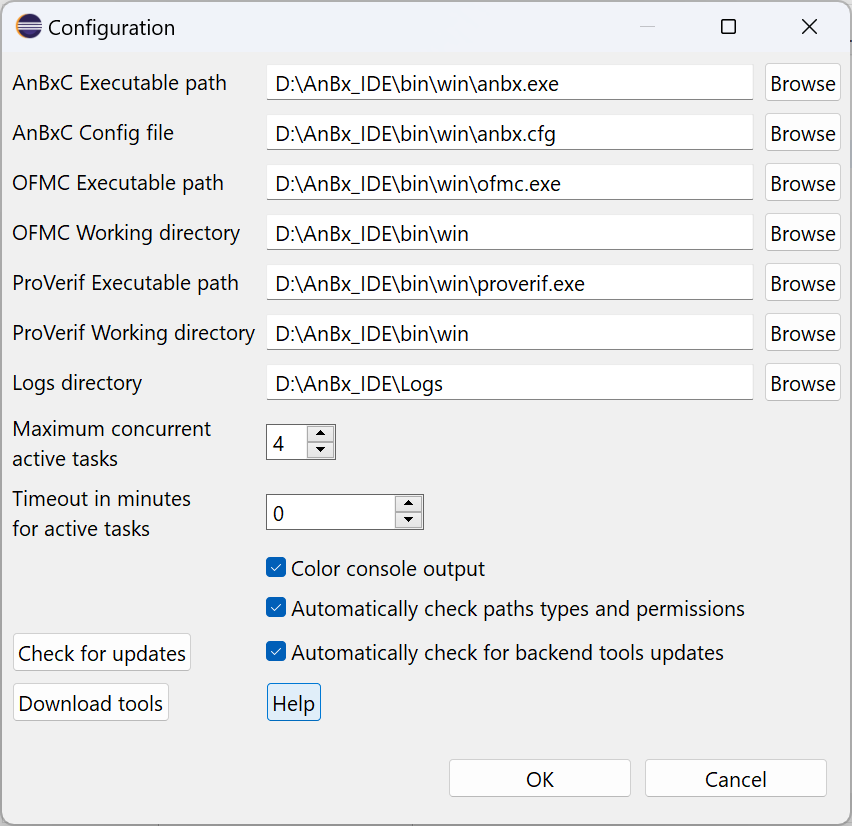}
\caption{\label{fig:settings}Configuration menu and window}
\end{figure}

\subsection{Paths for Tools and Logs}

The first entries in the configuration window (Figure \ref{fig:settings}) specify the paths for tool executables, configuration files, and working directories. These paths can be set using the file explorer or entered manually.

As described in Section \ref{sec:idecontribution}, each tool can log its output to a dedicated file within a folder defined in the configuration window. Subdirectories will be created for each tool, and files will be named based on the target protocol, the export option (if applicable), and a timestamp.

\begin{figure}[!h]
\centering
\includegraphics[scale=0.36]{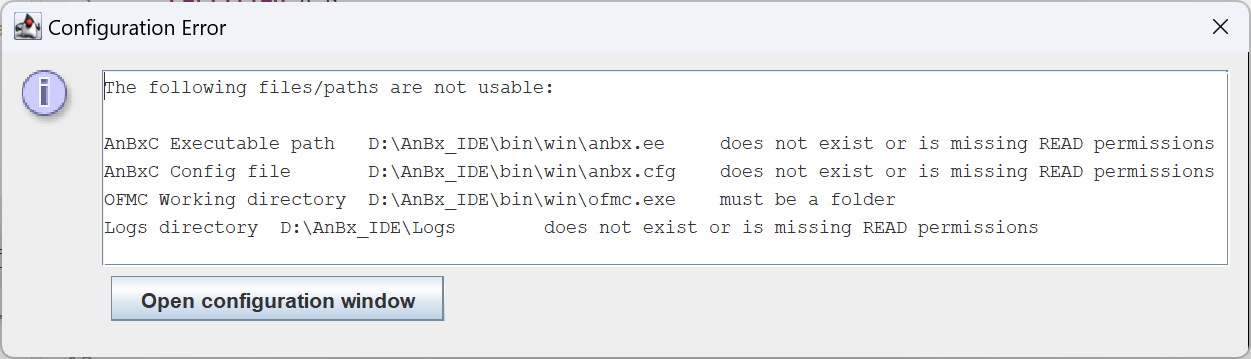}
\caption{\label{fig:pathscheck}Error messages for invalid paths}
\end{figure}

When the \textit{OK} button is pressed, the IDE verifies the existence and type of each path. If any issues are detected, an error message will be displayed in a popup window, as shown in Figure \ref{fig:pathscheck}.

\subsubsection{File System Permissions Checking}

By default, the IDE checks the file system to ensure that each path entered in the configuration window has sufficient permissions. Specifically:
\begin{itemize}
    \item Executables require \textbf{read} and \textbf{execute} permissions.
    \item Configuration files require \textbf{read} and \textbf{write} permissions.
    \item Folders require \textbf{read}, \textbf{write}, and \textbf{execute} permissions for traversal.
\end{itemize}

These checks help to prevent user confusion, especially when downloaded tools do not run correctly due to improper permissions. Since the checks are performed not only when running the tools but also at configuration time, they minimise disruptions caused by cross-tool interactions or logging errors.

In the unlikely case that these checks report false positives, they can be disabled in the configuration window.

\subsection{Links to Tools and Help}

\begin{figure}[!h]
\centering
\includegraphics[scale=0.4]{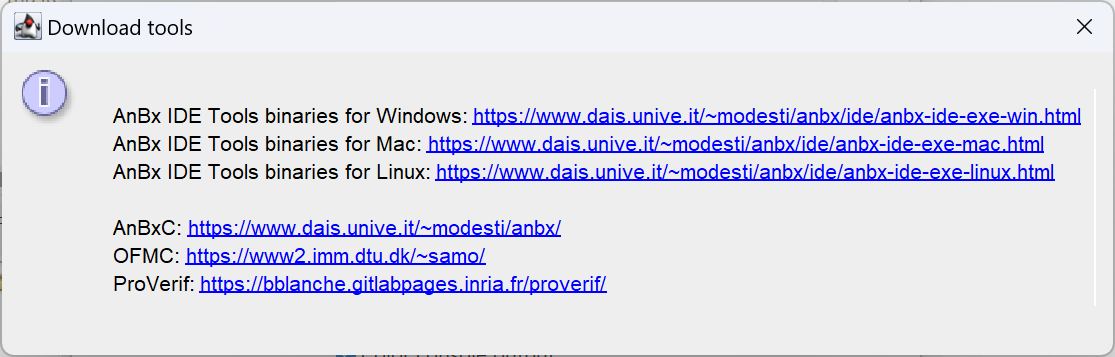}
\caption{\label{fig:downloadtools}Download tools window}
\end{figure}

The configuration window (Figure \ref{fig:settings}) includes buttons providing links to resources. The \textit{Download tool} button (Figure \ref{fig:downloadtools}) opens a dialog with links to binaries for Windows, Mac, and Linux, as well as the home pages of the tools.

\begin{figure}[!h]
\centering
\includegraphics[scale=0.5]{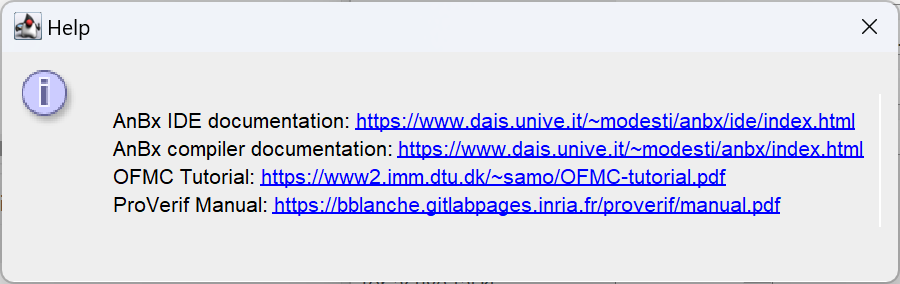}
\caption{\label{fig:help}Help window}
\end{figure}

The \textit{Help} button (Figure \ref{fig:help}) provides documentation links for the tools and the IDE itself, enabling users to get started quickly.

\subsection{Automatic or Manual Updates}

\begin{figure}[!h]
\centering
\includegraphics[scale=0.7]{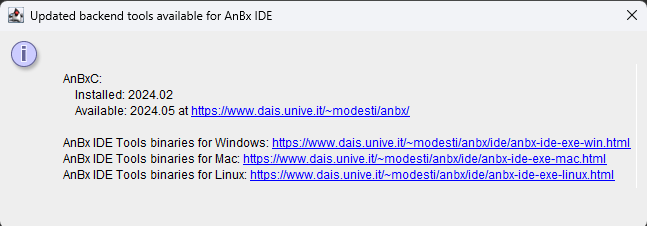}
\caption{\label{fig:Updates}Update window for back-end tools}
\end{figure}

The IDE tracks the currently available versions of AnBxC, OFMC, and ProVerif. When the paths to the relevant executables are set, the IDE checks if they correspond to the latest versions. If a newer, supported, version is available, a popup window will notify the user of the current version, the latest version, and provide relevant download links for each tool. An example is shown in Figure \ref{fig:Updates}.

In the configuration window, users can disable automatic update checks or trigger them manually using the \textit{Check for updates} button. This button will either display a message stating that no updates were found or open an update window, as in Figure \ref{fig:Updates}.

Automatic update checks occur once a week starting from the last check or whenever a tool's path is modified.

\section{Verifying Security Protocols}\label{app:verification}

This section describes how to verify security protocols using two tools: OFMC and ProVerif. The verification process ensures that the protocol specification meets its intended security goals.

\subsection{OFMC Protocol Verification}

\begin{figure}[!h]
	\centering
	\includegraphics[width=\textwidth]{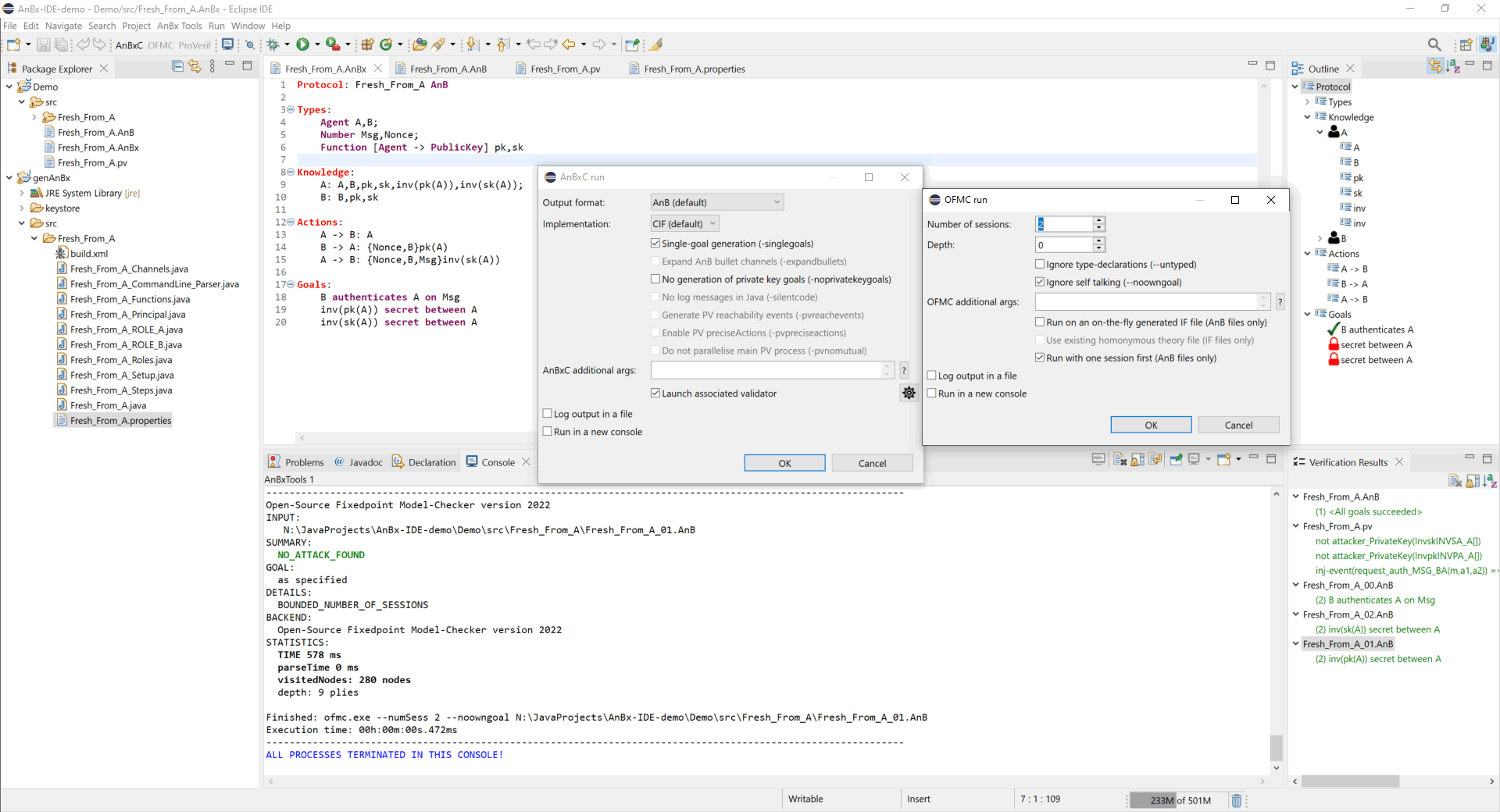}
	\caption{\label{fig:ofmc-verification}OFMC Protocol Verification}
\end{figure}

OFMC is used to verify protocols specified in \texttt{.AnBx}, \texttt{.AnB}, or \texttt{.IF} files. The process is shown in Figure \ref{fig:ofmc-verification}.

\subsubsection{Verifying \texttt{.AnBx} Files}

\begin{enumerate}
	\item Select the \texttt{.AnBx} file in Eclipse and click the \textit{AnBxC} button in the Eclipse toolbar.
	\item Choose the output format \textit{AnB (default)} and tick \textit{Launch associated validator}.
	\item (Only the first time) Configure OFMC:
	\begin{itemize}
		\item Click on the \includegraphics[width=4mm]{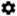} icon.
		\item Specify the number of sessions (e.g., 1, 2, 3, etc.).
		\item If the session number (\texttt{n}) is greater than 1, the tool will sequentially verify the protocol for 1 session and \textit{n} sessions.
		\item Click \textit{OK}.
	\end{itemize}
	\item The compiler will translate the file from \texttt{.AnBx} to \texttt{.AnB} and automatically run OFMC to verify the protocol against its specified security goals.
\end{enumerate}

\subsubsection{Verifying \texttt{.AnB} or \texttt{.IF} Files}
\begin{enumerate}
	\item Select the \texttt{.AnB} or \texttt{.IF} file in Eclipse and click the \texttt{OFMC} button in the Eclipse toolbar.
	\item Configure OFMC as follows:
	\begin{itemize}
		\item Specify the number of sessions (e.g., 1, 2, 3, etc.).
		\item If the session number (\textit{n}) is greater than 1, the tool will sequentially verify the protocol for 1 session and \textit{n} sessions.
	\end{itemize}
	\item Click \textit{OK} to begin the verification process.
\end{enumerate}

\subsection{ProVerif Protocol Verification}

\begin{figure}[!h]
	\centering
	\includegraphics[width=\textwidth]{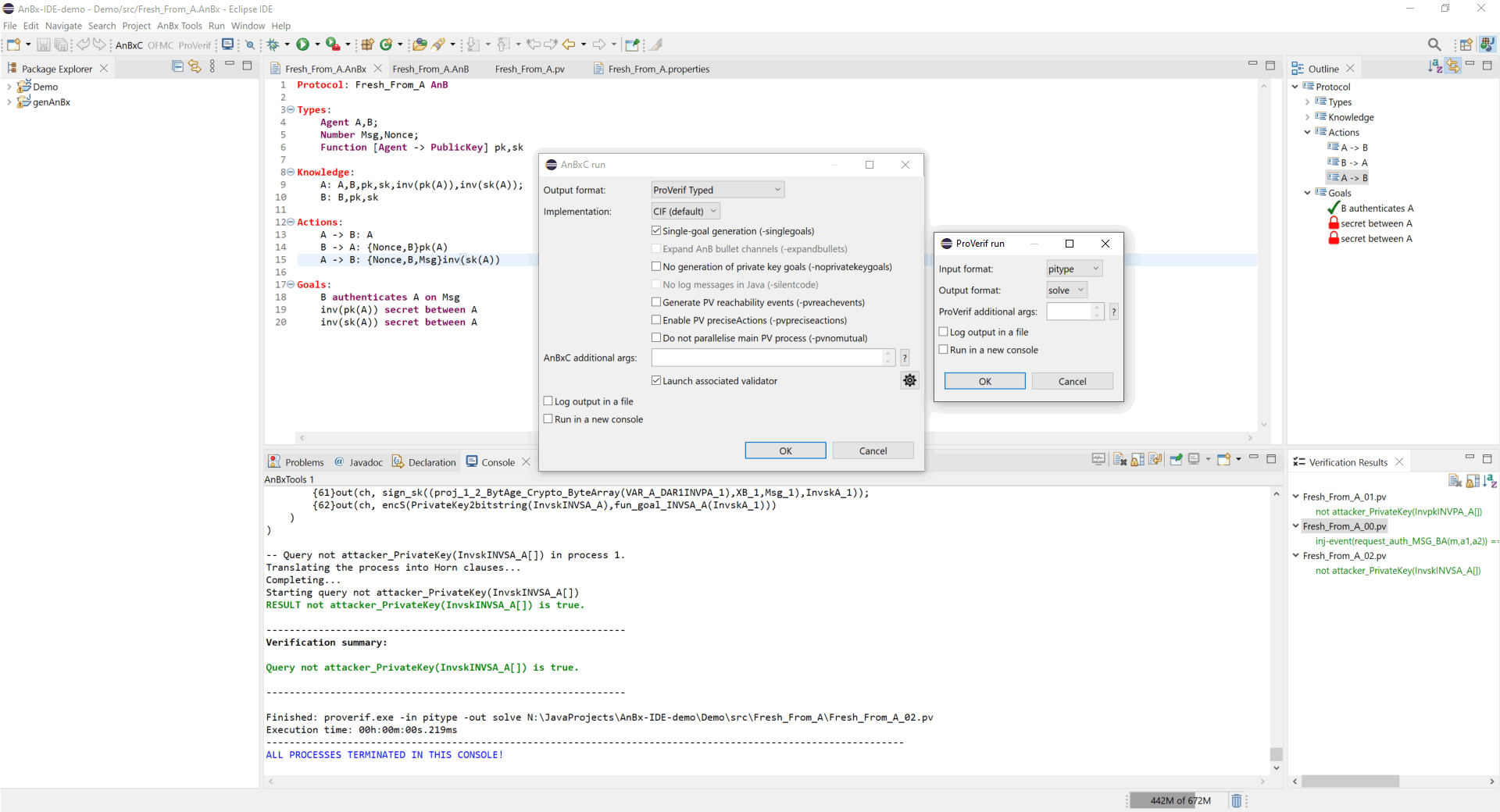}
	\caption{\label{fig:proverif-verification}ProVerif Protocol Verification}
\end{figure}

ProVerif is used to verify protocols specified in \texttt{.AnBx} or \texttt{.PV} files. It supports advanced verification of security goals using symbolic modelling. The process is shown in Figure \ref{fig:proverif-verification}.

\subsubsection{Verifying \texttt{.AnBx} Files}
\begin{enumerate}
	\item Select the \texttt{.AnBx} file in Eclipse and click the \textit{AnBxC} button in the Eclipse toolbar.
	\item Choose the output format \textit{ProVerif (default)} and tick \textit{Launch associated validator}.
	\item (Optional) Configure ProVerif:
	\begin{itemize}
		\item Click on the \includegraphics[width=4mm]{images/settings2.png} icon to set options for ProVerif.
		\item Options include \textit{pitype} (process type inference) and \textit{solve} (solver for symbolic traces).
	\end{itemize}
	\item Click \textit{OK}. The compiler will translate the file from \texttt{.AnBx} to \texttt{.PV} and automatically run ProVerif to verify the protocol against its specified security goals.
\end{enumerate}

\subsubsection{Verifying \texttt{.PV} Files}
\begin{enumerate}
	\item Select the \texttt{.PV} file in Eclipse and click the \textit{ProVerif} button in the Eclipse toolbar.
	\item Configure the verification options:
	\begin{itemize}
		\item Default options are \textit{pitype} and \textit{solve}.
	\end{itemize}
	\item Click \textit{OK} to run ProVerif and verify the protocol.
\end{enumerate}

\subsection{Notes on Verification}
\begin{itemize}
	\item Verification results are displayed in the Eclipse console or associated output window.
	\item If verification fails, inspect the error messages and ensure that the protocol and its configuration are correctly defined.
\end{itemize}

\section{Generating and Running Java Code with Eclipse}\label{app:java-code}

This section explains how to generate and run Java implementations of security protocols using the \anbx{} Compiler with \anbx{} IDE Eclipse plug-in. 

\subsection{Configuration in Eclipse}

To set up the Eclipse environment:
\begin{enumerate}
    \item Download and configure the Eclipse plug-in as described in \ref{app:getting-started}.
    \item Ensure the configuration file \texttt{anbxc.cfg} is correctly set up. Key parameters include:
    \begin{itemize}
        \item \texttt{pathstemplates}: Location of the template files (\texttt{.st}).
        \item \texttt{pathjavadest}: Path where the generated Java files will be stored.
    \end{itemize}
    \item Verify that \texttt{pathjavadest} points to the source folder of your Java project in Eclipse (e.g., \texttt{C:/JavaProjects/genAnBx/src/}).
\end{enumerate}

\subsection{Generating Java Code}

To generate Java code for an AnBx protocol file (e.g., \texttt{Fresh\_From\_A.AnBx}) using Eclipse:
\begin{enumerate}
    \item Open your Eclipse workspace.
    \item Click on the \textit{AnBxC} button in the toolbar.
    \item Select the desired output format (\textit{Java} for standard generation).
    \item Press \textit{OK}.
\end{enumerate}

The generated files will include Java source files and a corresponding \texttt{.properties} configuration file for the protocol.

\subsection{Setting Up and Running the Project in Eclipse}

To run the generated Java code in Eclipse:
\begin{enumerate}
    \item \textit{Create a New Java Project:}
    \begin{itemize}
        \item Navigate to \textit{File $\rightarrow$ New $\rightarrow$ Project $\rightarrow$ Java Project}.
        \item Specify a project name and location (e.g., \texttt{C:/JavaProjects/genAnBx}).
        \item Click \textit{Finish}.
    \end{itemize}
    \item \textit{Add the \texttt{AnBxJ.jar} Library:}
    \begin{itemize}
        \item Right-click on the project folder and select \textit{Properties}.
        \item Navigate to \textit{Java Build Path $\rightarrow$ Libraries}.
        \item If \texttt{AnBxJ.jar} is not listed, click \textit{Add External JARs...} and select the \texttt{AnBxJ.jar} file.
    \end{itemize}
    \item \textit{Ensure Proper Configuration:}
    \begin{itemize}
        \item Verify that the \texttt{keypath} in \texttt{anbxc.cfg} points to the correct keystore location.
        \item The default value for keystores is:
        \begin{verbatim}
# Paths
keypath = ../../keystore/
        \end{verbatim}
        \item Expand the keystore file into a folder at the same level as the \texttt{src} folder (e.g., \texttt{C:/JavaProjects/genAnBx/src/KeyStore}).
        \item Ensure that \texttt{pathjavadest} in \texttt{anbxc.cfg} is set to your project’s source folder (e.g., \texttt{C:/JavaProjects/genAnBx/src/}).
    \end{itemize}
    \item \textit{Generate and Import Code:}
    \begin{itemize}
        \item After generating Java code, The generated files will appear under the specified protocol folder (e.g., \texttt{src/Prot} for protocol \texttt{Prot}).
        \item  If you do not see the new files, refresh the folder to load them.
    \end{itemize}
    \item \textit{Run the Project:}
    \begin{itemize}
    	\item If \textit{Run associated validator} is enabled (see Figure \ref{fig:anbxcdialog}), the program will execute immediately after being built.
        \item Otherwise, use the \texttt{build.xml} ant file generated alongside the Java code.
        \item Right-click on the \texttt{build.xml} file and select \textit{Run As $\rightarrow$ Ant Build}.
    \end{itemize}
    \item If the application does not run as expected:
    \begin{itemize}
        \item Refresh the workspace and rebuild the project.
        \item Check the console output for details on errors or missing dependencies.
    \end{itemize}
\end{enumerate}

\section{Additional Features and Components}
Information about these features and components can be found in Section \ref{sec:idecontribution}:

\begin{itemize}
	\item  \ref{subsec:ideediting} Editing Security Protocol models in different specification languages
	\begin{itemize}
		\item \ref{subsec:ide-wizard} Getting started with wizards 
		\item \ref{subsec:ide-syntax} Syntax highlighting and outline	
		\item \ref{subsec:ide-formatting} Formatting 
		\item \ref{subsec:ide-autocomplete} Autocomplete and scoping 
		\item \ref{subsec:ide-validation} Validation: type, arity, and semantics checking with quickfixes 
	\end{itemize}
	\item \ref{subsec:ide-running} Running the verification tasks and generating code 
\begin{itemize}
	\item \ref{subsec:ide-configuration-dialogs} Configuration dialogs with options and help 
	\item \ref{subsec:ide-supporting-users} Supporting users in their workflow 
	\item \ref{subsec:ide-parallel-goals} Verifying single or multiple security goals in parallel 
	\item \ref{subsec:ide-java} Java Code Generation and Run of Dockerised applications 
	\item \ref{subsec:consoleoutput} Console output and logging 
	\item \ref{subsec:ideverifresultsview} Displaying verification results in an Eclipse view 
	\item \ref{subsec:taskmanager} Scheduling with priorities and Task manager  
	\item \ref{subsec:idetracereconstruction} Attack trace reconstruction 
\end{itemize}
	\item \ref{subsec:ide-environment} Environment customisation and other functionalities 
\end{itemize}

\end{document}